\documentclass[journal]{IEEEtran}

\usepackage[backend=biber,style=ieee]{biblatex}
\usepackage{graphicx}
\graphicspath{ {./images/} }
\DeclareGraphicsExtensions{.png}

\usepackage{amsmath}
\usepackage{algorithm}
\usepackage{algorithmic}
\usepackage{listings} 

\usepackage{array}

\usepackage{url}

\begin{document}
\title{Advancing fNIRS Neuroimaging through Synthetic Data Generation and Machine Learning Applications}

\author{Eitan~Waks,~\IEEEmembership{Johns~Hopkins~University,}
}

\maketitle

\begin{abstract}
This study presents an integrated approach for advancing functional Near-Infrared Spectroscopy (fNIRS) neuroimaging through the synthesis of data and application of machine learning models. By addressing the scarcity of high-quality neuroimaging datasets, this work harnesses Monte Carlo simulations and parametric head models to generate a comprehensive synthetic dataset, reflecting a wide spectrum of conditions. We developed a containerized environment employing Docker and Xarray for standardized and reproducible data analysis, facilitating meaningful comparisons across different signal processing modalities. Additionally, a cloud-based infrastructure is established for scalable data generation and processing, enhancing the accessibility and quality of neuroimaging data. The combination of synthetic data generation with machine learning techniques holds promise for improving the accuracy, efficiency, and applicability of fNIRS tomography, potentially revolutionizing diagnostics and treatment strategies for neurological conditions. The methodologies and infrastructure developed herein set new standards in data simulation and analysis, paving the way for future research in neuroimaging and the broader biomedical engineering field.
\end{abstract}


\IEEEpeerreviewmaketitle

\section{Background: History of functional near-infrared spectroscopy}
\IEEEPARstart{T}{he} development of optical methods for assessing changes in the optical properties of brain tissue began with Glenn Millikan's invention of the muscle oximeter in the forties \cite{RefWorks:RefID:6-chance1991optical}. Frans Jöbsis founded in vivo near-infrared spectroscopy (NIRS), utilizing the transparency of brain tissue in the NIR range to detect hemoglobin oxygenation noninvasively \cite{RefWorks:RefID:7-jöbsis1977noninvasive}. Marco Ferrari used prototype NIRS instruments to measure changes in brain oxygenation in experimental animal models and human adults. From 1980 to 1995, several companies collaborated with universities to develop NIRS prototypes \cite{RefWorks:RefID:9-ferrari2012review}.

Functional imaging is the assessment of physiological changes associated with brain activity. Functional MRI (fMRI) based on blood oxygenation level dependent (BOLD) imaging was first suggested in 1990 \cite{RefWorks:RefID:8-ogawa1990brain} and was followed by the discovery of human functional near-infrared spectroscopy (fNIRS) in 1992 by by Chance, Kato, Hoshi, and Villringer \cite{RefWorks:RefID:9-ferrari2012review}. fNIRS detects changes in the optical properties of the cortex and provides maps or images of specific areas. The increase in oxygenated hemoglobin and the decrease in deoxygenated hemoglobin reflect an increase in local blood flow and volume due to neurovascular coupling.

fNIRS is based on human tissues being relatively transparent to light in the NIR spectral window (650-1000 nm) and that NIR light can penetrate tissues due to scattering being more probable than absorption. Hemoglobin is the main chromophore that attenuates NIR light in tissue, and its absorption spectrum depends on its level of oxygenation. In the NIR spectral window light absorption increases as a function of frequency for oxygenated hemoglobin and decreases for deoxygenated hemoglobin. The absorption characteristics are equal at approximately 810 nm. This is the isosbestic point. The Beer-Lambert Law relates the attenuation of light to the concentration of absorbing species in a medium, e.g. oxygenated hemoglobin and deoxygenated hemoglobin.

In turbid media such as biological tissues, light scattering is a significant factor that must be considered. The Modified Beer-Lambert Law is a general form of the Beer-Lambert Law that takes into account the effects of light scattering in addition to light absorption by a medium. The Modified Beer-Lambert Law extends the Beer-Lambert Law to include the effects of light scattering by introducing a scattering coefficient in addition to the absorption coefficient.

The cortical hemodynamic response to brain activity is characterized by hyper oxygenation. This is due to the metabolic demands of neuronal activity. Temporally, the response is characterized by an initial dip on the order of milliseconds followed by an increase in oxygenation which occurs on the order of seconds. Increases typically initiate two seconds after stimulus. A return to resting state concentration levels occurs approximately 16 seconds after stimulus \cite{RefWorks:RefID:10-friston1998nonlinear}.

fNIRS systems are comprised of a power source connected to a controller, instructions from the controller are relayed through a preamplifier to light sources. The choice of light sources, such as laser diodes (LDs) and light emitting diodes (LEDs), depends on the specific requirements of the study. LDs have narrow spectral peaks, emit coherent light, and are suitable for fiber optic coupling, but pose a potential danger to the eyes. LEDs emit incoherent light, have a wide bandwidth, and can be easily adjusted in intensity. The selection of wavelengths affect the quality of measurement data. Light sources with sharply peaked radiation spectra, such as monochromatic light, are desirable for fNIRS measurements at several discrete wavelengths. Selecting optimal wavelengths depends on several variables, including the number of wavelengths used, the number and type of chromophores considered, the model of the background medium, and the mathematical approach to solve the optimization problem \cite{RefWorks:RefID:11-scholkmann2014review}. fNIRS devices can be bifurcated to those who use 2 discrete wavelengths and those who use 3 or more wavelengths. For two-wavelength fNIRS devices, the optimum wavelength in combination with 830 nanometers (nm) should be < 780 nm \cite{RefWorks:RefID:11-scholkmann2014review}. Furthermore, the optimal wavelength pair for a two-wavelength fNIRS device appears to be 704 $\pm$ 7 + 887 $\pm$ 12 nm \cite{RefWorks:RefID:12-correia2010identification}. There are two options to transfer light from source to medium of interest, the first by placing sources and detectors directly on the skin, the second by guiding light through optical fibers to probes on the head. Direct placement has minimal light coupling losses but potential hazards, while fiber optic transmission allows for more flexible probe design but has added weight and reduced mobility. 

The scattered photons which exit the scalp are detected by photodiodes. The two most common types of photo diodes used in fNIRS Systems are photodiodes (PD) and avalanche photodiodes (APD) \cite{RefWorks:RefID:11-scholkmann2014review}. Charge couple devices (CCD's) can also be used to detect photons and have an advantage for detecting broadband signals as well as providing spatial information. For optimal performance these detectors should be power stabilized and amplified using little noise amplifiers. Silicon photomultipliers (SiPMs) maybe used as detectors for fNIRS instrumentation, which enables measurements at larger source-detector separations (SDS) while affording small, lightweight, and modular probes \cite{RefWorks:RefID:13-zimmermann2013silicon}. SiPMs enable measurements at an SDS up to 50 millimeters (mm), which is 67\% more than the commonly used SDS of 30 mm \cite{RefWorks:RefID:9-ferrari2012review}, have large internal amplification, and allow for simple and miniaturizable data readout, thus making SiPMs desirable as photo detectors.

There are three techniques implemented by fNIRS systems; continuous wave (CW), time domain (TD), and frequency domain (FD) fNIRS systems. CW-fNIRS and fd-fNIRS systems emit a constant (non-pulsating) signal whereas td-fNIRS systems emit a high frequency pulsating signal. CW-fNIRS measure attenuation and cannot provide absolute values of hemoglobin concentration. Contrastly, fd-fNIRS measure both attenuation and phase differences. The added information from the modulating light source in the form of a phase shift can provide additional information with higher resolution than that of CW technique. TD-fNIRS measure changes in pulse shape and are extremely fast, on the order of picoseconds, however equipment is very expensive and sensitive to artifacts when compared to the other two techniques.



\section{Methods}

 The data in this study were recorded under the approval of an IRB approved by the Johns Hopkins University Investigation Review Board.
 
The datasets are stored in the network common data form (NetCDF) file format \cite{RefWorks:RefID:18-2023netcdf:}. NetCDF is a binary file format used to store scientific data and metadata in a self-describing form. It is a flexible format that can be used to store a variety of data types, including arrays, charts, and tables. NetCDF files can contain multiple data variables, each with its own dimensions, attributes, and data. The dimensions of a variable describe the size of the data array, while the attributes provide additional information about the variable, such as its name, units, and data type.

Xarray \cite{RefWorks:RefID:20-hoyer2017xarray:} is used to load the datasets. Xarray is a Python library that provides support for labeled, multi-dimensional arrays (also known as N-dimensional arrays or NDArrays) and integrates with a range of other scientific Python packages. Xarray provides a high-level interface for working with NetCDF files.
It is built on top of NumPy and Pandas libraries, providing many of the same features.

Xarray has two core data structures, \texttt{DataArray} and \texttt{Dataset}, both of which are fundamentally N-dimensional. \texttt{DataArray} is a labeled, N-dimensional array. It is an N-D generalization of a \texttt{pandas.Series}. The name \texttt{DataArray} itself is borrowed from Fernando Perez’s datarray project \cite{RefWorks:RefID:22-perez2023bids/datarray:}, which prototyped a similar data structure. \texttt{Dataset} is a multi-dimensional, in-memory array database. It is a \texttt{dict}-like container of \texttt{DataArray} objects aligned along any number of shared dimensions, and serves a similar purpose in xarray to the \texttt{pandas.DataFrame}.

Coordinates are ancillary variables stored for DataArray and Dataset objects in the \texttt{coords} attribute. Coordinates indicate constant/fixed/independent quantities, unlike the varying/measured/dependent quantities that belong in data. Xarray does interpret and persist coordinates in operations that transform xarray objects. There are two types of coordinates in xarray:
\begin{itemize}
    \item dimension coordinates - one dimensional coordinates used for label based indexing and alignment with a name equal to their dimension.
    \item non-dimension coordinates - multidimensional variables useful for indexing or plotting which contain coordinate data, but are not a dimension coordinate. Xarray does not make any direct use of the values associated with them, and are not used for alignment or automatic indexing, nor are they required to match when doing arithmetic.
\end{itemize}

The \texttt{load\_dataset} method needs the engine parameter set to: \texttt{h5netcdf}. h5netcdf is a Python library that provides support for reading and writing data to and from NetCDF files using the HDF5 format. h5netcdf is built on top of the netCDF4-python library and the h5py library, which provide low-level access to the NetCDF and HDF5 formats, respectively. It provides a high-level, easy-to-use interface for working with NetCDF files, and is particularly useful for reading and writing large datasets, as it can efficiently handle data that does not fit into memory. Xarray’s lazy loading of remote or on-disk datasets is often but not always desirable. Before performing computationally intense operations, it is often a good idea to load a Dataset (or DataArray) entirely into memory by invoking the \texttt{Dataset.load()} method.

\section{Data Analysis Environment}
Installing and configuring software and its dependencies on multiple machines is time-consuming and error-prone. Containers are a lightweight and portable way to package an application and all its dependencies, including libraries and system tools, into a single, self-contained unit thereby simplifying and enhancing reproducibility and data provenance. Containers are isolated from each other and from the host system, allowing you to run multiple applications on a single host without them interfering with each other. Docker \cite{RefWorks:RefID:21-merkel2014docker:} is an open-source platform for creating, deploying, and managing software applications using containers. Docker containers can be run on any machine with the Docker runtime installed. 

A Docker container is an instance of a Docker image, which is a lightweight, standalone, executable package that includes everything needed to run an application, including code, runtime, system tools, libraries, and settings. Docker images are built from a Dockerfile, which is a script that specifies the application's dependencies and how they should be configured.

For this project, a Dockerfile was used to build an image for a Jupyter notebook with the necessary dependencies for fNIRS. The image is built upon the minimal-notebook image from the Jupyter Docker Stacks \cite{RefWorks:RefID:16-2023jupyter/docker-stacks:}. Jupyter Docker Stacks are a set of pre-configured Docker images containing Jupyter applications and interactive computing tools. The minimal-notebook image is the most basic image, containing only the minimal packages necessary to run a Jupyter notebook server.

The Dockerfile then installs the additional dependencies specified in the requirements.txt file using the mamba package manager \cite{RefWorks:RefID:17-2023mamba-org/mamba:}. Mamba is a fast drop-in replacement for the conda package manager, which is included in the minimal-notebook image. It is designed to be used in place of conda for running pipelines, including inside of Docker containers.

The requirements.txt include the following packages:
\begin{itemize}
    \item python: The Python interpreter
    \item numpy: NumPy is a library for numerical computing in Python.
    \item scipy: SciPy is a library that builds on top of NumPy to provide additional functionality for scientific computing.
    \item xarray: Xarray is a library for working with labeled multi-dimensional arrays in Python.
    \item matplotlib: Matplotlib is a library for creating static, animated, and interactive visualizations in Python.
    \item seaborn: Seaborn is a library for creating statistical visualizations in Python.
    \item hdf5: A library for reading and writing HDF5 files.
    \item h5py: h5py is a library for working with HDF5 files in Python.
    \item ipykernel: ipykernel is a library that provides a kernel for Jupyter notebooks.
    \item jupyterlab: The JupyterLab interface for Jupyter notebooks.
    \item ipympl: ipympl is a library that provides interactive plotting for Jupyter notebooks using the matplotlib library.
    \item ipywidgets: ipywidgets is a library that provides interactive widgets for Jupyter notebooks.
    \item iprogress: iprogress is a library that provides a progress bar for Jupyter notebooks.
    \item tqdm: tqdm is a library that provides a progress bar for loops and other iterative processes in Python.
    \item pathlib: pathlib is a library for working with file paths in a platform-independent way.
    \item tabulate: tabulate is a library that provides a simple and flexible way to print tables in Python.
    \item mne: MNE is a library for analyzing neurophysiological data in Python.
    \item mne-nirs: MNE-NIRS is an extension to the MNE library that provides tools for analyzing fNIRS data.
\end{itemize}

Overall, these packages provide a wide range of functionality for scientific computing, data analysis, and visualization in Python. They are widely used in academia and industry for a variety of applications, including neuroscience.


Automating containerization can further simplify deployment and scaling processes, increase consistency and reliability, and reduce the risk of errors and downtime, ultimately improving overall system performance and stability. Docker Compose is a widely used tool in the field of containerization that enables the easy deployment of containers. It allows users to define the required services, dependencies, networks, and volumes in a single file, which can be used to create and start all the required containers with a single command (docker-compose up). Volumes are useful for persisting data between service runs. Docker Compose also provides support for orchestration and deployment in clustered environments.

In this study the Docker Compose file defines a single service, \texttt{fnirs}, which builds and runs a Docker image for fNIRS data analysis. The image is built using the Dockerfile in the Docker directory, and is based on the \texttt{fnirs:latest} image. The \texttt{fnirs} service is exposed on port 8888 and is configured with two volumes:

\begin{enumerate}
    \item \texttt{./data:/home/jovyan/work/data:ro}: This volume mounts the \texttt{./data} directory on the host machine to the \texttt{/home/jovyan/work/data} directory in the container, and sets it as read-only.
    \item \texttt{./notebooks:/home/jovyan/work/notebooks:rw}: This volume mounts the \texttt{./notebooks} directory on the host machine to the \texttt{/home/jovyan/work/notebooks} directory in the container, and sets it as read-write.
\end{enumerate}

Interacting with the \texttt{fnirs} service is done through a web browser. Any changes to the files are saved locally and may subsequently pushed to a cloud-based repository.

\section{Datasets}
\subsection{Dataset 01}
Dataset 01 is comprised of three experiment runs. The name for each run corresponds to the date and time it was recorded e.g., 20220825T1416 is a run recorded August 25, 2022 at 2:16 PM. Each run (variable) has a 3 dimensional tuple index, collectively referred to as "\texttt{dimensions}", consisting of \texttt{time}, \texttt{src}, and \texttt{det} components. The \texttt{dimensions} correspond to the an individual run value's time, source and detector coupling. The value data type is \texttt{complex128} (aka \texttt{numpy.cdouble(real, imag)}). The values are complex to capture both the magnitude and phase components of the fd-fNIRS system.

There are 13 coordinates associated with this dataset: \texttt{det}, \texttt{dposx}, \texttt{dposy}, \texttt{dposz}, \texttt{lbd}, \texttt{sposx}, \texttt{sposy}, \texttt{sposz}, \texttt{src}, \texttt{time}, \texttt{NN}, \texttt{r2d}, and \texttt{r3d}. Of these 3 are dimensional and 10 non-dimensional, 10 are one dimensional and 3 multidimensional. 

Values are uniquely identified by time and source detector coupling. Accordingly we must uniquely define each of these parameters. The coordinate \texttt{det} and \texttt{src} corresponds to detector and source identifiers, respectively. There are 32 detectors and 64 sources in the data. Detector and source identifiers are labeled as integers ranging from 0 to 31 and 63 respectively, i.e., 0, 1, 2, ..., (31/63). Both detectors and sources each have 32 physical sources, however each source emits 2 different wavelengths therefore we have twice the amount of sources (64) in the data. Sources i and i+32 are co-located. Sources 0-31 emit light at a frequency of 690 nm whereas sources 32-63 emit light at 850 nm. Sources are switched on and off sequentially within a time encoding block, the duration of which is 10 ms, per wavelength. The coordinate \texttt{time} corresponds to time in seconds and spans a range of 4.99856 - 844.75684, with consistent intervals at 0.49985612 seconds. Accordingly, we have 1680 data points for \texttt{time}.

Since we are interested in exploring spatiotemporal phenomena we must know the spatial information (in three dimensions) for each detector and source. Coordinates \texttt{dposx}, \texttt{dposy}, \texttt{dposz}, \texttt{sposx}, \texttt{sposy}, and \texttt{sposz} refer to the distance in mm between the origin and a detector/source along the x, y, or z axis. The origin, in relation to the skull, is defined as the coincidence of the following planes:

\begin{itemize}
    \item The plane parallel to the mid sagittal plane 9.25 mm lateral of detector 0 when the strip of detectors is embedded within the coronal plane.
    \item The plane parallel to the transverse plane 9.25 mm inferior the first detector
    \item The coronal plane coinciding with the occipital pole.
\end{itemize}

Meaningful data was provided for the \texttt{x} and \texttt{y} positions yet values for the \texttt{z} positions were all 0. Since we do not have accurate measurements of the subject's head, we must estimate the \texttt{z} values. To do so a function was written based upon the paper "Centiles for adult head circumference" \cite{RefWorks:RefID:23-bushby1992centiles}. Bushby et al. present a model for adult human head circumference as a function of height, sex, and head circumference centile. The transverse plane cross sectional geometry of the human head is not circular in nature therefore estimating \texttt{z} values based on the Bushby et al. model can only provide an approximation of the true values.

Our subject is a 185 cm tall male. The model predicts a mean head circumference of 54.8 cm. On January 5, 2023 the subject measured his head circumference with a tape ruler. The result was 58 cm, approximately to the 75$^{th}$ percentile in the Bushby et al. model.

\begin{figure}[!t]
    \centering
    \includegraphics[width=\columnwidth]{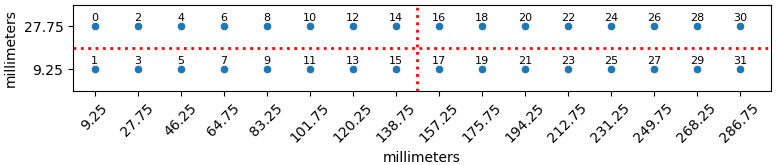}
    \caption{Detector positions displayed in 2-D. The detectors are embedded within a strip that is placed on the rear of the subject's head. The occipital pole is at the coincidence of the symmetry lines which are represented as red dotted lines.}
    \label{fig:det_pos_2D}
\end{figure}

\begin{figure}[!t]
    \centering
    \includegraphics[width=\columnwidth]{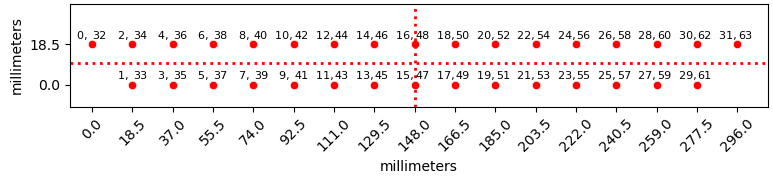}
    \caption{Source positions on the strip placed on the subjects head, displayed in 2-D. The red dotted lines represent symmetry lines. The coincidence of the symmetry lines align with the occipital pole.}
    \label{fig:src_pos_2D}
\end{figure}

\begin{figure}[!t]
    \centering
    \includegraphics[width=\columnwidth]{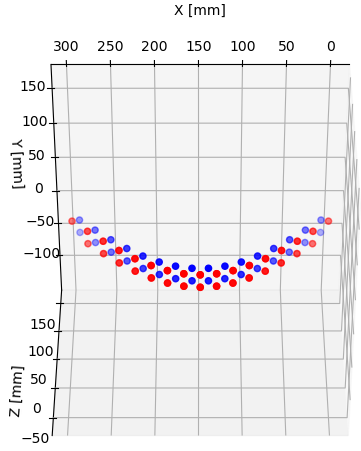}
    \caption{Sensor and detector positions displayed in 3-D. Detectors are represented as blue dots. Sensors are represented as red dots. The \texttt{z} values were estimated using Bushby et al.'s model for estimating adult head circumference. The X plane is the plane parallel to the mid sagittal plane 9.25 mm lateral to detector 0 when the strip of detectors is embedded within the coronal plane. The Y plane is the plane parallel to the transverse plane 9.25 mm inferior of detector 0. The Z plane is the plane parallel to the coronal plane and coincident with the occipital pole.}
    \label{fig:src_det_pos}
\end{figure}

To estimate \texttt{dposz} values we project the two dimension positional values for sources and detectors onto a cylinder defined by a circumference equal to the subject head circumference. Radius is a function of circumference: 
$$ r = \frac{C}{2\pi} $$
The angle from the occipital point to a source/detector as a function of radius and curvature length is:  
$$ \theta = \frac{l}{r} $$
The estimated distance of a source/detector from the occipital point along the \texttt{z} axis is:  
$$ d = r(1-\cos\theta) $$
\texttt{dposz} and \texttt{sposz} values were updated accordingly. Figure \ref{fig:det_pos_2D} and figure \ref{fig:src_pos_2D} show the positions of detectors and sources in two dimensions, respectively. Sources and detectors interweave equidistantly. Figure \ref{fig:src_det_pos} shows the relative position of sources and detectors in three dimensions. 

The typical trajectory of the NIR light emitted is described as "banana-shaped" function depicting the probability density of photon path-lengths \cite{RefWorks:RefID:24-van1990effect}. The distance between source and detector "lengthen" or "shorten" the distance between the edges of the banana function hence affecting the probability of detected photon penetration depth. Increased source detector separation (SDS) increase the probability of detecting photons which have penetrated deeper into the brain. Accordingly, it is desirable to rank SDS. Coordinate \texttt{NN} is a 64 x 32 element array depicting no nearest neighbor pair rank between source and detector. 

\begin{figure}[!t]
    \centering
    \includegraphics[width=\columnwidth]{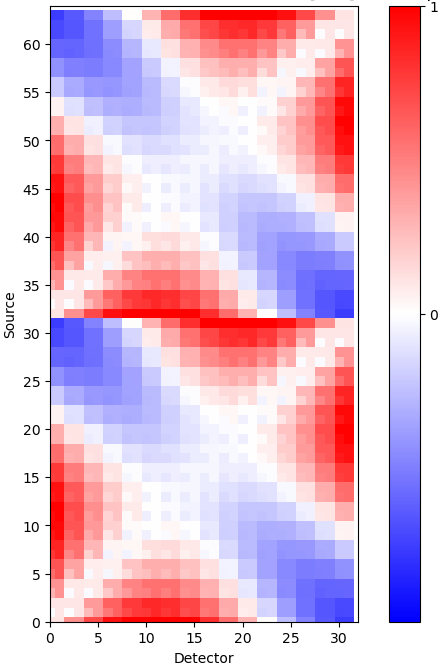}
    \caption{A red-blue heat map depicting the difference in 2-D vs 3-D Euclidean (L2 Norm) distance between sources and detectors. This is a normalized heat map where red is is positive displacement and blue is negative. The differences increase as NN increases.}
    \label{fig:2D_v_3D_euc_dist}
\end{figure}

The difference in two dimensional versus three-dimensional Euclidean distance between the sources and detectors may have a significant impact on diffused optical tomography (DOT) calculations. The difference becomes more significant others \texttt{NN} values increase. Figure \ref{fig:2D_v_3D_euc_dist} shows the differences. The differences are likely to be greater if the banana functions were used as opposed to the Euclidean distance.

\begin{figure}[!t]
    \centering
    \includegraphics[width=\columnwidth]{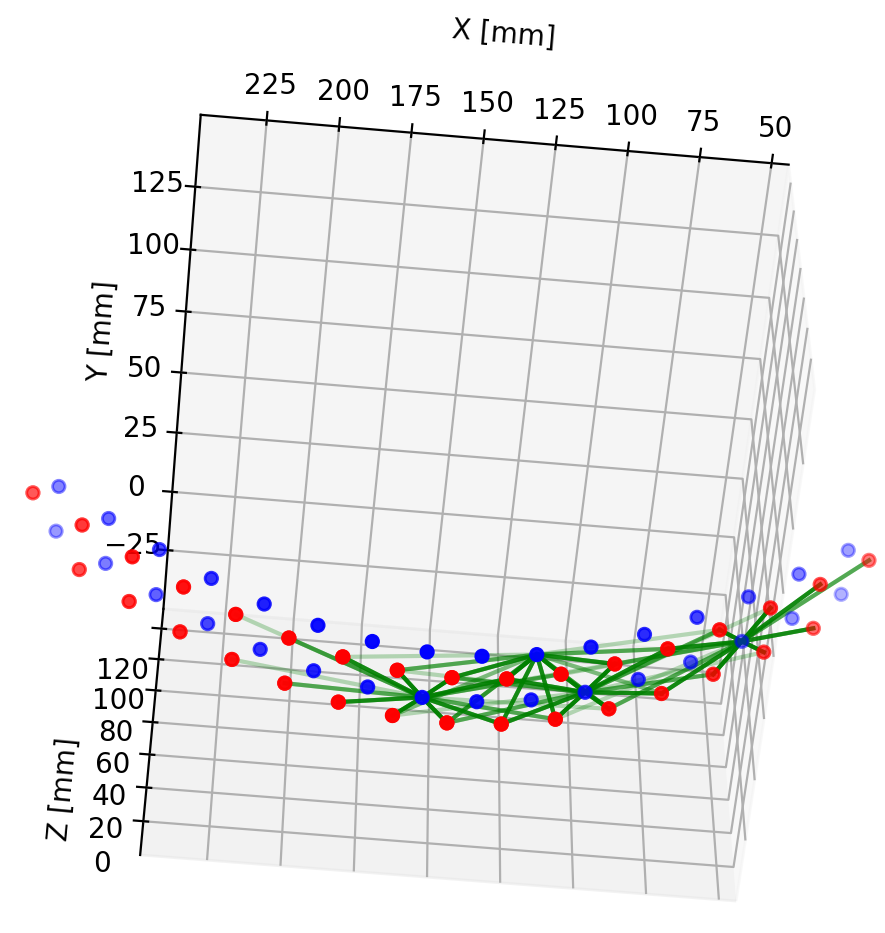}
    \caption{The probability of a detector sensing a photon from a particular source is a negative function of SDS. The complexity and probability of photon paths is schematically represented. Overlapping paths may be used for blind tomography. Schematic photon propagation for source detector pairs with a maximum \texttt{NN} (SDS) value of 7 for detectors 5, 11, 12, 17 is displayed as green lines. The green lines are L2 norm distances where saturation decreases with increasing \texttt{NN} values.}
    \label{fig:det_src_NN}
\end{figure}

To gain more intuition as to the complexity of photon propagation with relation to SDS a function was written which visualizes the Euclidean distance between source in detector pairs. The saturation of the line connecting the source and detector is a function of separation rank i.e. \texttt{NN}. Figure \ref{fig:det_src_NN} visualizes the SDS as a function of source detector pair.

\begin{figure}[!t]
    \centering
    \includegraphics[width=\columnwidth]{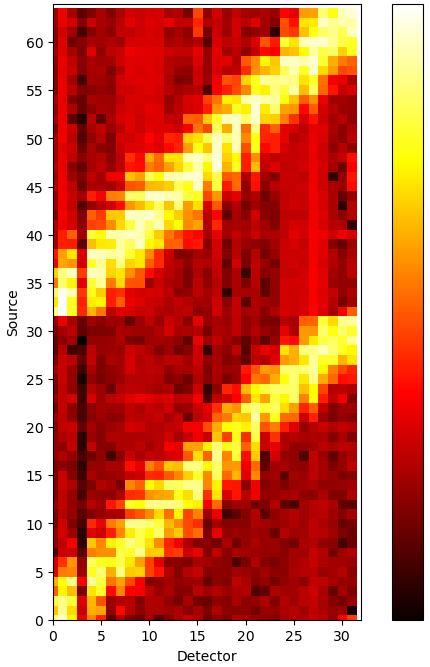}
    \caption{A heatmap of the mean of magnitude across time for each source detector pair is displayed. Detectors are resources which exhibit consistently low or high averages across time we are likely to be faulty. In this example detector 3 is likely faulty. Detectors 24 - 27 seem to be exceptionally well coupled with sources 32-63.}
    \label{fig:det_src_coupling}
\end{figure}

In theory, the range of values for source detector pairs with the same SDS should be equal. In practice, this dynamic range and signal-to-noise ratio are affected by a myriad of factors, including the degree of source detector coupling. Detecting faulty sources or detectors is important so they may be excluded from further analysis. One possibility for examining coupling strength between source and detector is comparing the mean of the magnitude (data) across time with the average value for source detector pairs with the same NN value, for each pair. Visualizing the values in the form of the heat map as in Figure \ref{fig:det_src_coupling} maybe an important tool for detection of coupling strength. Detector 3 in Figure \ref{fig:det_src_coupling} appears to be badly coupled to the scalp or faulty. An in operable detector or source would have the same value for each source detector pair and presumably be viewed as a solid black line (vertical for detector, horizontal for source).

\subsection{Dataset 02}
 Semantic representation refers to the process of representing meaning or content of language in a structured format that can be processed by computers or other intelligent systems. The goal of semantic representation is to capture the meaning of language in a way that can be used to reason about, understand, and manipulate the content of natural language text. 

Mitchell et al. \cite{RefWorks:RefID:25-mitchell2008predicting} developed a model for predicting nouns based on fMRI imaging. Both fNIRS and fMRI imaging measure hemodynamic responses. Using fMRI has several disadvantages when compared to fNIRS. Some of these disadvantages are limited availability, high financial burden, limited temporal resolution, unnatural testing environment, and susceptibility to motion artifacts. To date, there is no model for transforming from fNIRS space to semantic space. Dataset 02 is oriented towards exploring the relationship between semantic representation and neurovascular coupling data obtained using fNIRS.

The dataset has one variable, \texttt{human218}, corresponding to a single run. The dataset's dimensions is a 2-D tuple consisting of \texttt{noun}, having 60 components, and \texttt{semantic\_feature} having 218 components. The value data type is \texttt{float64}. There are 3 coordinates all of which are one dimensional. \texttt{noun}, \texttt{feature\_desc}, and \texttt{group\_membership}. The latter two are non-dimensional coordinates.

The trial was to think of an object. The objects were chosen from Mitchell's \cite{RefWorks:RefID:25-mitchell2008predicting} list of 60 individual nouns: 'bear' 'cat' 'cow' 'dog' 'horse' 'arm' 'eye' 'foot' 'hand' 'leg', 'apartment' 'barn' 'church' 'house' 'igloo' 'arch' 'chimney' 'closet', 'door' 'window' 'coat' 'dress' 'pants' 'shirt' 'skirt' 'bed' 'chair', 'desk' 'dresser' 'table' 'ant' 'bee' 'beetle' 'butterfly' 'fly' 'bottle', 'cup' 'glass' 'knife' 'spoon' 'bell' 'key' 'refrigerator' 'telephone', 'watch' 'chisel' 'hammer' 'pliers' 'saw' 'screwdriver' 'carrot' 'celery', 'corn' 'lettuce' 'tomato' 'airplane' 'bicycle' 'car' 'train' 'truck'.

These nouns can be grouped in several different ways, for example:
\begin{itemize}
    \item Grouping by category type (Table \ref{table_grouping_by_category_type})
    \item Grouping by category function (Table \ref{table_grouping_by_category_function})
    \item Grouping by category appearance (Table \ref{table_grouping_by_category_appearance})
\end{itemize}

\begin{table}[!t]
\renewcommand{\arraystretch}{1.3}
\caption{Grouping by category type}
\label{table_grouping_by_category_type}
\centering
\begin{tabular}{|m{2.5cm}||m{5cm}|}
\hline
Animals & bear, cat, cow, dog, horse\\
\hline
Body parts & arm, eye, foot, hand, leg\\
\hline
Buildings & apartment, barn, church, house, igloo\\
\hline
Building parts & arch, chimney, closet, door, window\\
\hline
Clothing & coat, dress, pants, shirt, skirt\\
\hline
Furniture & bed, chair, desk, dresser, table\\
\hline
Insects & ant, bee, beetle, butterfly, fly\\
\hline
Kitchenware & bottle, cup, glass, knives, spoon\\
\hline
Machines/Devices & bell, key, refrigerator, telephone, watch\\
\hline
Tools & chisel, hammer, pliers, saw, screwdriver\\
\hline
Vegetables & carrot, celery, corn, lettuce, tomato\\
\hline
Vehicles & airplane, bicycle, car, train, truck\\
\hline
\end{tabular}
\end{table}

\begin{table}[!t]
\renewcommand{\arraystretch}{1.3}
\caption{Grouping by category function}
\label{table_grouping_by_category_function}
\centering
\begin{tabular}{|m{2.5cm}||m{5cm}|}
\hline
Living things & bear, cat, cow, dog, horse, ant, bee, butterfly, fly\\
\hline
Buildings and Building parts & apartment, barn, church, house, igloo, arch, chimney, closet, door, window\\
\hline
Clothing & coat, dress, pants, shirt, skirt\\
\hline
Furniture & bed, chair, desk, dresser, table\\
\hline
Kitchenware & bottle, cup, glass, knives, spoon\\
\hline
Machines/Devices & bell, key, refrigerator, telephone, watch\\
\hline
Tools & chisel, hammer, pliers, saw, screwdriver\\
\hline
Food & carrot, celery, corn, lettuce, tomato\\
\hline
Transportation & airplane, bicycle, car, train, truck\\
\hline
\end{tabular}
\end{table}

\begin{table}[!t]
\renewcommand{\arraystretch}{1.3}
\caption{Grouping by category appearance}
\label{table_grouping_by_category_appearance}
\centering
\begin{tabular}{|m{2.5cm}||m{5cm}|}
\hline
Large Objects & airplane, barn, church, house, igloo, train, truck\\
\hline
Small Objects & ant, bee, beetle, bell, key, knife, spoon, watch\\
\hline
Clothing & coat, dress, pants, shirt, skirt\\
\hline
Furniture & bed, chair, desk, dresser, table\\
\hline
Food & carrot, celery, corn, lettuce, tomato\\
\hline
Insects & ant, bee, butterfly, fly\\
\hline
Animals & bear, cat, cow, dog, horse\\
\hline
Body parts & arm, eye, foot, hand, leg\\
\hline
Building parts & arch, chimney, closet, door, window\\
\hline
Machines/Devices & refrigerator, telephone\\
\hline
Tools & chisel, hammer, pliers, saw, screwdriver\\
\hline
\end{tabular}
\end{table}

Feature description is a list of questions used to describe an object. The full list is provided in Appendix \ref{appendix_questions}.

 The list can be used as a set of prompts or guidelines to help classify objects or things into different categories. The categories include animals, body parts, buildings, building parts, clothing, furniture, insects, kitchen items, man-made objects, tools, vegetables/plants, vehicles, persons and colors and textures.

The list is divided into subcategories and the questions are tailored to each subcategory for example:

For the animal subcategory, it has questions such as:
\begin{itemize}
\item Does it have a tail?
\item Does it have legs?
\item Does it have four legs?
\item Does it have paws?
\item Does it have claws?
\item Does it have hooves?
\item Does it have wings?
\item Does it have feathers?
\item Does it have some sort of fur/hair?
\item Does it have scales?
\item Does it have a shell?
\item Does it have a spine?
\end{itemize}

For the color subcategory, it has questions such as:
\begin{itemize}
\item Is it colorful?
\item Does it change color?
\item Is one more than one colored?
\item Is it always the same color(s)?
\item Is it white?
\item Is it red?
\item Is it orange?
\item Is it yellow?
\item Is it green?
\item Is it blue?
\item Is it silver?
\item Is it flesh-colored?
\end{itemize}

By answering yes or no to these questions, you can determine which category an object belongs to. For example, if you're trying to classify a lion, you would answer "yes" to the questions such as:
\begin{itemize}
\item Does it have a tail?
\item Does it have four legs?
\item Does it have paws?
\item Does it have a face?
\item Does it have a mane?
\item Does it have some sort of fur/hair?
\item Does it have a spine?
\item Does it have a pointed/sharp tail?
\end{itemize}

Similarly, if you're trying to classify a couch, you would answer "yes" to the questions such as:
\begin{itemize}
\item Is it furniture?
\item Does it have a flat/straight top?
\item Does it have flat/straight sides?
\item Is taller than it is wide/long?
\item Is it long?
\item Is it symmetrical?
\item Is it soft?
\end{itemize}

It's important to note that the list of questions provided is not exhaustive and may not be able to classify every single object, but it serves as a guide to help classify objects into the given categories. Additionally, the categorization may not be mutually exclusive, an object can fall under multiple categories, for example a car can be considered as man-made object, vehicle and have colors.

\section{Advancing fNIRS Tomography Through Machine Learning and Synthetic Data Generation}

Development of accurate and robust analytical methods is crucial for advancing our understanding of brain function and improving the applicability of fNIRS in clinical and research settings. Given the complexities associated with fNIRS data, there is a growing interest in developing advanced analytical tools to enhance data interpretation. The use of machine learning techniques holds the potential for efficiently analyzing, classifying, and predicting physiological states from fNIRS signals \cite{RefWorks:RefID:31-benerradi2023benchmarking}. However, the effectiveness of such techniques is directly tied to the availability of extensive, high-quality datasets, which are often challenging to procure in neuroimaging due to ethical, logistical, and financial constraints.

To address this challenge, we generated synthetic data through Monte Carlo simulations of photon propagation in tissue. This approach enables the creation of comprehensive and varied datasets, capable of embodying a range of conditions. The synthetic dataset serves as a controlled framework for establishing reliable ground truth, essential for the future development and validation of machine learning models. This process can particularly enhance the performance and generalizability of supervised learning models, which rely on labeled data, while also providing opportunities for unsupervised learning models to identify patterns without predefined labels, thereby potentially enhancing their predictive performance when applied to actual fNIRS data, such as Dataset 01 and Dataset 02.

In the following sections, we outline the processes involved in synthetic data generation and preparation for future machine learning applications in fNIRS tomography. We elaborate on the methodology for constructing parametric head models, simulating photon migration patterns, and developing a scalable cloud environment designed to orchestrate extensive simulations efficiently. This groundwork is critical for enabling the subsequent application of machine learning techniques to fNIRS data analysis, aiming to provide insights into brain activity and improve the technique's utility in various settings.

\section{Synthetic Data Generation for fNIRS}
Synthetic data generation is a viable approach in addressing the scarcity of comprehensive datasets for fNIRS analysis. Utilizing Monte Carlo simulations for photon propagation in tissue, synthetic data generation enables the creation of extensive, controlled datasets. These datasets can encompass a wide range of conditions, establishing a reliable ground truth for future analytical model development and validation. This approach is particularly beneficial in overcoming the challenges associated with acquiring large-scale, diverse neuroimaging data.

In the field of medical physics, Monte Carlo simulations are utilized extensively for generating large datasets, especially in particle transport simulations. These simulations are crucial for understanding complex phenomena such as dose distribution in radiotherapy and imaging. The efficiency of these computational methods has been greatly enhanced by modern computing infrastructures, such as clusters and GPUs, facilitating the generation of results in an acceptable time frame. This approach aligns with efforts in high-energy physics, where Monte Carlo simulations are employed to improve understanding and accelerate computations \cite{RefWorks:RefID:32-sarrut2021artificial}.

Current methods for generating fNIRS data primarily rely on real experimental data and time series analysis. Experimental data, derived from actual fNIRS studies, offers insights into brain activity under specific conditions but lacks anatomical specifics crucial for tomography \cite{RefWorks:RefID:33-gabrieli2021fnirs-qc:}. Time series datasets, while valuable for understanding dynamic brain responses, often do not provide the anatomical detail necessary for in-depth analysis \cite{RefWorks:RefID:34-gemignani2018improving}. These methods, though informative, limit the scope of machine learning applications due to the absence of a defined ground truth and anatomical context.

In this work, we employ an approach to synthetic data generation for fNIRS using Monte Carlo simulations, which differ significantly from conventional methods by producing snapshots in time rather than full time series. This unique strategy allows us to create extensive, anatomically accurate datasets, providing the detailed ground truth necessary for advanced machine learning applications. Unlike previous methods, our approach facilitates labeled machine learning on a large scale by leveraging known anatomical models. This novel methodology hopes to advance neuroimaging research by overcoming traditional data collection hurdles.

This section delves into synthetic data creation, strategic employment of parametric models, the integration of Monte Carlo simulations for realistic data synthesis, and an overview of brain atlases for accurate spatial representation, thus setting the stage for advanced analytical developments in fNIRS tomography.

\subsection{Strategies for Synthetic Data Creation}

Various methodologies exist for generating synthetic data, each with its unique strengths and application areas. Notable methods include:

\begin{itemize}
    \item Generative Adversarial Networks (GANs): Utilizes two neural networks, a generator and a discriminator, to create data indistinguishable from real datasets.
    \item Agent-based Modeling: Simulates the actions and interactions of autonomous agents to generate complex phenomena.
    \item Bootstrap Resampling: Involves creating new datasets by randomly sampling with replacement from an existing dataset.
    \item Parametric Simulation: Assumes a specific distribution for data and generates samples from this distribution.
    \item Monte Carlo Simulation: Employs random sampling to understand variability and uncertainty in systems.
\end{itemize}

Each method caters to different aspects of synthetic data needs, depending on the complexity and type of data required.

\subsection{Fundamentals and Applications of Monte Carlo Methods in fNIRS}

Monte Carlo simulations are a statistical technique used for understanding complex systems. This approach is particularly useful in the context of fNIRS, where it helps model the propagation of light through biological tissues. 

The principle of Monte Carlo simulations lies in the random sampling to approximate numerical results for deterministic problems. This allows for the exploration of all possible outcomes and assesses the likelihood of different outcomes occurring.

Mathematically, if we consider a random variable $X$ with a probability density function $f(x)$, the expected value of $X$ is defined as:

\begin{equation}
    E[X] = \int x f(x) \, dx
\end{equation}

However, directly computing such integrals can be challenging, especially for complex functions. Monte Carlo methods approach this by estimating the expected value as:

\begin{equation}
    E[X] \approx \frac{1}{N} \sum_{i=1}^{N} x_i
\end{equation}

where $x_i$ are independent identically distributed samples from $f(x)$.

Furthermore, the law of large numbers supports Monte Carlo simulations, stating that as the number of samples increases, the average of these samples converges to the expected value.

The accuracy of Monte Carlo estimates can be expressed through variance:

\begin{equation}
    \text{Var}\left(\frac{1}{N} \sum_{i=1}^{N} X_i \right) = \frac{\sigma^2}{N}
\end{equation}

where $\sigma^2$ is the variance of $X$. The standard error, $\sigma/\sqrt{N}$, measures the estimate's potential fluctuation from the true value.

The application of Monte Carlo simulations expanded with the invention of digital computing, reducing the computational cost associated with random sampling. The rise of distributed computing and parallel processing, alongside advancements in computational speed and algorithms, has made Monte Carlo simulations a standard, robust, and efficient tool in scientific research and beyond, effectively enabling the handling of complex, uncertain systems like photon propagation in fNIRS studies.

\subsection{Mathematics of Photon Propagation and Monte Carlo eXtreme (MCX)}

Monte Carlo simulations have proven invaluable in modeling photon migration through turbid media such as human tissues. These simulations are conducted by launching a vast number of photons and tracking their paths through the medium, effectively simulating each photon's random walk. This approach estimates the probability distribution of photon paths and energy deposition without directly solving complex differential equations. This method is particularly advantageous in turbid, low-scattering media, establishing it as the gold standard in bio-optical imaging applications like functional brain imaging \cite{RefWorks:RefID:35-fang2009monte}.

Photon propagation in biological tissues is generally governed by the Radiative Transfer Equation (RTE), a fundamental equation describing light transport in scattering and absorbing media. The RTE is expressed as:

\begin{equation}
\begin{split}
&\frac{1}{c} \frac{\partial L(\mathbf{r}, \mathbf{s}, t)}{\partial t} + \mathbf{s} \cdot \nabla L(\mathbf{r}, \mathbf{s}, t) + [\mu_a(\mathbf{r}) + \mu_s(\mathbf{r})] L(\mathbf{r}, \mathbf{s}, t) = \\ 
&\qquad \mu_s(\mathbf{r}) \int_{4\pi} p(\mathbf{r}, \mathbf{s}', \mathbf{s}) L(\mathbf{r}, \mathbf{s}', t) d\Omega' + S(\mathbf{r}, \mathbf{s}, t),
\end{split}
\end{equation}

where \(L(\mathbf{r}, \mathbf{s}, t)\) is the radiance, \(\mu_a(\mathbf{r})\) and \(\mu_s(\mathbf{r})\) are the absorption and scattering coefficients, respectively, \(p(\mathbf{r}, \mathbf{s}', \mathbf{s})\) is the phase function, and \(S(\mathbf{r}, \mathbf{s}, t)\) is the source term.

Solving the RTE directly in complex media like human tissue is challenging, which is why Monte Carlo methods, known for their simplicity and adaptability to low-scattering media, are often preferred.

The key parameters governing the photon migration within the Monte Carlo simulations include:

Scattering Coefficient (\(\mu_s\)): This parameter dictates the frequency of photon scattering events within the medium. It is defined as the number of scattering events per unit path length. A higher \(\mu_s\) means photons scatter more frequently. The scattering coefficient plays a role in determining the path and intensity distribution of light in tissue:
\begin{equation}
    \mu_s = \frac{1}{\text{mean free path of scattering}}
\end{equation}
    
Absorption Coefficient (\(\mu_a\)): This measures the rate at which photons are absorbed by the medium. Similar to the scattering coefficient, it is expressed as the number of absorption events per unit path length. The absorption coefficient directly impacts the intensity of light as it propagates through the tissue:
\begin{equation}
    \mu_a = \frac{1}{\text{mean free path of absorption}}
\end{equation}
    
Anisotropy Factor (\(g\)): The anisotropy factor describes the scattering angle dependency, typically ranging from -1 (perfect backscatter) to 1 (perfect forward scatter). This factor influences the phase function and, subsequently, the scattering profile of photons within the medium:
\begin{equation}
    g = \langle \cos \theta \rangle
\end{equation}
where \(\theta\) is the scattering angle.

Refractive Index (\(n\)): The refractive index determines how photons are refracted at the boundaries between different media. It affects the boundary conditions and the Fresnel reflections and transmissions at interfaces:
\begin{equation}
    n = \frac{c_{medium}}{c_{vacuum}}
\end{equation}
where \(c_{medium}\) and \(c_{vacuum}\) are the speeds of light in the medium and vacuum, respectively.

Additionally, the simulation accounts for the photon weight (\(W\)), which decreases with each scattering or absorption event, reflecting the loss of light intensity due to these interactions:

\begin{equation}
W_{new} = W_{old} \times (1 - \frac{\mu_a}{\mu_a + \mu_s})
\end{equation}

Time-resolved measurements and boundary conditions are also integral to accurately modeling photon migration. The time-dependent nature of photon movement can be expressed using the temporal point spread function (TPSF), which is a measure of the photon density as a function of time:

\begin{equation}
TPSF(t) = \frac{1}{4\pi D t} \exp\left(-\frac{r^2}{4Dt} - \mu_a t\right)
\end{equation}
where \(D\) is the diffusion coefficient, and \(r\) is the distance from the light source.

Monte Carlo eXtreme (MCX) \cite{RefWorks:RefID:35-fang2009monte} represents a significant advancement in this field, utilizing parallel computing on graphics processing units (GPUs) to enhance the speed of photon migration simulations substantially. MCX enables the simultaneous simulation of millions of photons, offering a significant speed increase compared to traditional CPU-based Monte Carlo methods. It supports both mesh-based and parametric modeling; the former provides high anatomical fidelity suitable for localized studies, while the latter offers a computationally less demanding approach for broader, generalized investigations.

\subsection{Brain Atlases in Monte Carlo Simulations: A Mesh-Based Approach}
In order to conduct Monte Carlo simulations for the study of light propagation in biological tissues, particularly the brain, it is necessary to know the optical properties of various brain components. Determining these properties requires defining the spatial domain through which photons will travel. Brain atlases are instrumental in mapping the complex anatomical and functional regions of the brain.

Historically, brain atlases were developed based on the post-mortem examination of human brains, exemplified by Brodmann atlas, which demarcated 52 distinct regions of the cerebral cortex based on variations in cytoarchitecture \cite{RefWorks:RefID:36-zilles2010centenary}. Advancements in medical imaging technologies, such as MRI and CT scans have enhanced the resolution and diversity of brain images available for study. This has facilitated the development of more detailed atlases, based on living subjects, thereby enabling the generation of atlases that better represent the variability found across different populations.

Anatomical variations between individuals and among different demographic groups have been observed. These variations provide incentive for generating atlases tailored towards specific populations. Notably, atlases that concentrate on the brain's vasculature are increasingly relevant in the context of fNIRS, where understanding the intricacies of cerebral blood flow is of prime importance.

Most atlases focus solely on brain tissue, overlooking non-brain structures such as the scalp and skull. These elements are necessary for Monte Carlo simulations of photon migration. There is a growing need for incorporating these non-brain structures in atlases. 

The creation of brain atlases follows a structured pipeline. The process, inspired by the LONI pipeline \cite{RefWorks:RefID:37-mandal2012structural}. This method begins with MRI data acquisition, providing the raw anatomical information for atlas creation. This data is pre-processed adjusting the raw MRI data for better quality and uniformity. Preprocessing steps include stripping to remove non-relevant tissue and normalizing the intensity of MRI images for consistency. Subsequently, the pre-processed data is segmented into different tissue types e.g., gray matter, white matter, cerebrospinal fluid, etc. The individual MRI scans are aligned and registered to a common standard space, ensuring consistency across different subjects. By averaging the registered MRI data, a composite image is created that represents the typical anatomy of the study population. This averaged data forms the basis of the brain atlas. The final step involves transforming the averaged MRI data into a three-dimensional mesh.

Many brain atlases exist, each with distinct characteristics. The Talairach and Tournoux Atlas (1988) set a historical precedent, offering a stereotaxic framework based on a single post-mortem brain, significantly impacted neurosurgical guidance and functional brain mapping \cite{RefWorks:RefID:38-talairach1988co-planar}.

The Montreal Neurological Institute (MNI) contributed with the MNI-305 \cite{RefWorks:RefID:39-evans19933d} (1995) and the MNI-152 \cite{RefWorks:RefID:41-diedrichsen2009probabilistic} (2001) templates, which are constructed from multiple MRI scans to represent a more average brain structure.

The Colin-27 (1998) template\cite{RefWorks:RefID:40-holmes1998enhancement}, derived from MRI scans of a single individual, presents a higher resolution alternative, enabling more detailed studies. Meanwhile, the ICBM-452 \cite{RefWorks:RefID:42-shattuck2008construction} (2003) template expands on this by averaging data from an extensive cohort, providing a more comprehensive representation of brain anatomy.

Cultural and demographic specificity in brain structure has led to the development of regional templates like the Korean Brain Template \cite{RefWorks:RefID:43-lee2005development} and the French Brain Template \cite{RefWorks:RefID:44-lalys2010construction}, addressing the need for more diverse anatomical benchmarks in neuroscience research.

Modern 7T MRI imaging has enabled the creation of specialized atlases such as the BigBrain project \cite{RefWorks:RefID:46-amunts2013bigbrain:}, offering unprecedented detail, and the Braincharter and VENAT \cite{RefWorks:RefID:47-huck2019high} atlases, focusing on cerebrovascular structures. These tools are of particular interest for fNIRS modeling, providing detailed maps of brain vasculature crucial for understanding and simulating light propagation in brain tissues.

\begin{table}[t]
    \caption{Chronological list of notable brain atlases with unique features.}
    \label{tab:brain_atlases}
    \centering
    \begin{tabular}{|l|c|p{3.6cm}|}
        \hline
        \textbf{Year} & \textbf{Atlas} & \textbf{Unique Feature} \\
        \hline
        \hline
        1988 & Talairach and Tournoux & Basis for neurosurgery \\
        \hline
        1995 & MNI-305 & Early average brain template \\
        \hline
        1998 & Colin-27 & Single-subject high detail \\
        \hline
        2001 & MNI-152 & Widely used modern standard \\
        \hline
        2003 & ICBM-452 & Broad cohort representation \\
        \hline
        2005 & Korean Brain & Reflects Korean anatomical variation \\
        \hline
        2009 & French Brain & High-resolution, French male \\
        \hline
        2010 & Chinese Brain & Highlights Chinese dimensions \\
        \hline
        2013 & BigBrain & Near-cellular ultrahigh resolution \\
        \hline
        2018 & Braincharter Cerebrovascular & Cerebral arteries and veins \\
        \hline
        2019 & VENAT & High-res venous vasculature \\
        \hline
        2022 & PAVI & High-res pial artery mapping \\
        \hline
    \end{tabular}
\end{table}

While brain atlases are invaluable for mesh-based Monte Carlo simulations due to their detailed anatomical mapping, they present drawbacks. The complexity involved in setting up these simulations, especially considering the configurations and parameters required, can be daunting. Furthermore, typical brain atlases lack comprehensive inclusion of non-brain tissues such as the scalp and skull, which are integral to bio-fidelic Monte Carlo simulations in fNIRS studies. This omission render many of these atlases inappropriate for simulations of light propagation in biological tissues, without extensive post processing for adding these tissues.

\section{Methodology}

\subsection{Parametric Head Models: Simplifying Complexity}

Given the inherent complexities associated with mesh-based simulations, particularly those that require detailed brain atlases, parametric models serve as a compelling alternative for initial stages of research or for establishing a proof of concept. The motivation for utilizing parametric models stems from their simplicity and flexibility. Unlike mesh-based models that rely on detailed anatomical structures, parametric models use geometric shapes and predefined parameters to approximate the human head and its internal structures. This simplification significantly reduces the setup time and computational resources required, enabling rapid testing and iteration of different scenarios.

The parametric head model used in our simulations represents a simplified yet effective approach, approximating the human head's complex anatomy with concentric spheres, each corresponding to different head tissues. This model strikes a balance between maintaining essential biological characteristics and computational efficiency. The adoption of a simplified geometry stems from the need to perform a large amount of computational simulations. Simplified parametric models benefit from reduced computational cost. While this approach may not capture all anatomical nuances, it is valuable in enabling broad, exploratory research within practical timeframes and resources.

Our parametric head model consists of four concentric spheres. The first sphere represents the scalp, followed by spherical representations of the skull, cerebrospinal fluid (CSF), and the brain. Within the brain, additional geometries are embedded representing blood vessels.

\begin{table}[t]
    \caption{Thickness of different head tissues used in the parametric model.}
    \label{tab:tissue_thickness}
    \centering
    \begin{tabular}{|l|c|}
        \hline
        \textbf{Tissue} & \textbf{Thickness (mm)} \\
        \hline
        \hline
        Scalp & 8 \\
        \hline
        Skull & 5 \\
        \hline
        CSF & 2.5 \\
        \hline
    \end{tabular}
\end{table}

Scalp thickness has been found to decrease with age, especially over the temporo-parietal area. A strong inverse correlation between age and scalp thickness exists, with measurements showing a decrease from a mean of 8 mm in the third decade of life to 5 mm by the ninth decade \cite{RefWorks:RefID:49-ungar2018age‐dependent}. In this study we will assume that scalp thickness in the occipital region and the temporo-parietal area are similar.

Penetration and scattering of light within the cerebral domain are significantly influenced by the skull's thickness and its inherent optical properties. The human skull, characterized by an inner and outer cortical table interspersed by the diploë, showcases thickness variations across its expanse and among individuals, with noticeable disparities based on age and gender. Drawing upon the findings from Lillie et al. (2016) \cite{RefWorks:RefID:50-lillie2016evaluation}, where evaluations were conducted using computed tomography scans to assess skull cortical thickness changes with age and gender, our parametric head model adopts a standardized thickness value of 5 mm. 

The meninges play a crucial role in protecting the brain, serving as a set of three layered membranes that encapsulate it. These layers, from the outermost to the innermost, are the dura mater, arachnoid mater, and pia mater. Particularly of interest is the subarachnoid space, sandwiched between the arachnoid and the pia mater. This space is filled with cerebrospinal fluid (CSF) and is responsible for the majority of the meninges' volume. Due to the significant volume occupied by the CSF in the subarachnoid space, it becomes a rational choice for parameterization in our head model. Such a representation allows for a more streamlined and anatomically representative model that efficiently captures the physiological significance of the meninges. In determining the appropriate thickness for the CSF parameter, we have settled on a value of 2.5 mm. This choice is grounded in the empirical findings presented by Parisa Saboori and Ali Sadegh in their detailed examination of the brain's subarachnoid trabeculae \cite{RefWorks:RefID:51-saboori2015histology}. 

\begin{figure}[!t]
    \centering
    \includegraphics[width=\columnwidth]{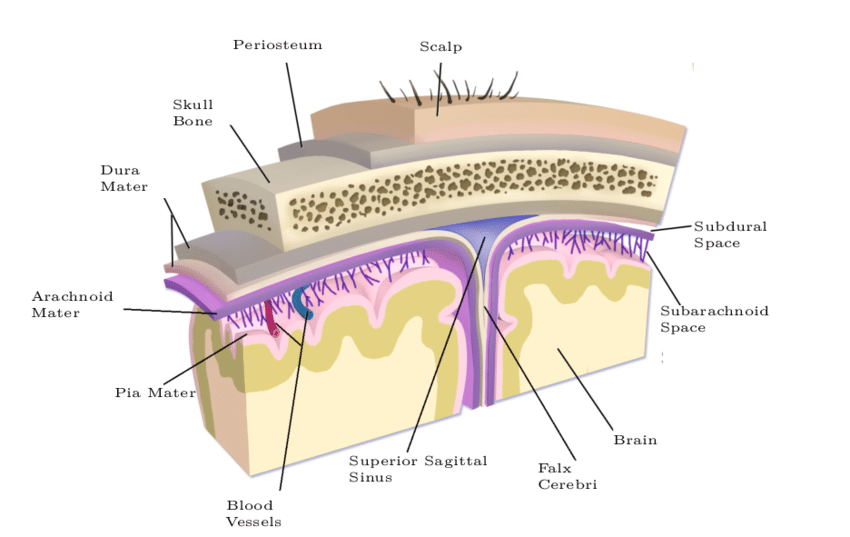}
    \caption{Anatomy of the human head showing the layers constituting the scalp, skull, meninges, and the brain. The illustration details the structural relationships between the periosteum, scalp, skull bone, dura mater, arachnoid mater, pia mater, superior sagittal sinus, falx cerebri, subdural space, subarachnoid space, and blood vessels. Adapted from De Kegel, D. (2018). Tissue-Level Tolerance Criteria for Crash-Related Head Injuries: a Combined Experimental and Numerical Approach \cite{RefWorks:RefID:55-de2018tissue-level}.}
    \label{fig:human_head_anatomy}
\end{figure}

\subsection{Optical Properties}
Following the establishment of the parametric head model, attention must be directed towards defining the optical properties of the different tissues within this simplified framework. Namely the absorption coefficient (\(\mu_a\)), the scattering coefficient (\(\mu_s\)), the anisotropy factor (\(g\)), and the refractive index (\(n\))—determine the interaction of light within the head's tissues.

Due to the head's complexity and variance among individuals, precise experimental data on the optical properties for adult human tissues remain limited. Historically, many studies utilized homogeneous models, leading to potential inaccuracies when applied to the head's diverse anatomical structures. The heterogeneity in reported values, stemming from different experimental methodologies and lack of standardization, complicates the establishment of universally accepted parameters. Jacques's review \cite{RefWorks:RefID:52-farina2015in-vivo} underscores this variability, noting that reported values for the brain's reduced scattering coefficient can vary significantly.

In this study, the optical properties were selected based on available literature, aiming to maintain consistency across the two wavelengths primarily used in our device, 690 nm and 850 nm. Given the lack of single-source comprehensive data, we rely on the closest available studies, adapting values from Tremblay et al. \cite{RefWorks:RefID:53-tremblay2018comparison} for general tissue properties and from Bosschaart et al. \cite{RefWorks:RefID:54-bosschaart2014literature} for blood-specific properties. This approach, while acknowledging potential discrepancies, provides a standardized foundation for our simulations, assuming minor differences in tissue properties between the slightly varied wavelengths of 830 nm and 850 nm are negligible. The optical properties values for the relevant tissues at 830 nm and 690 nm are displayed in tables \ref{tab:optical_properties_830} and \ref{tab:optical_properties_690}, respectively.

\begin{table}[t]
    \caption{Optical properties of different head tissues at 830 nm wavelength.}
    \label{tab:optical_properties_830}
    \centering
    \begin{tabular}{|l|c|c|c|c|}
        \hline
        \textbf{Tissue} & \textbf{$\mu_a$ (mm$^{-1}$)} & \textbf{$\mu_s$ (mm$^{-1}$)} & \textbf{g} & \textbf{n} \\
        \hline
        \hline
        Scalp  & 0.0191 & 0.66 & 0.9 & 1.4 \\
        \hline
        Skull  & 0.0136 & 0.86 & 0.9 & 1.4 \\
        \hline
        CSF    & 0.0260 & 0.01 & 0.9 & 1.4 \\
        \hline
        Brain  & 0.0186 & 1.11 & 0.9 & 1.4 \\
        \hline
        Blood  & 0.46   & 75.06 & 0.9835 & 1.33 \\
        \hline
    \end{tabular}
\end{table}

\begin{table}[t]
    \caption{Optical properties of different head tissues at 690 nm wavelength.}
    \label{tab:optical_properties_690}
    \centering
    \begin{tabular}{|l|c|c|c|c|}
        \hline
        \textbf{Tissue} & \textbf{$\mu_a$ (mm$^{-1}$)} & \textbf{$\mu_s$ (mm$^{-1}$)} & \textbf{g} & \textbf{n} \\
        \hline
        \hline
        Scalp  & 0.0159 & 0.80 & 0.9 & 1.4 \\
        \hline
        Skull  & 0.0101 & 1.00 & 0.9 & 1.4 \\
        \hline
        CSF    & 0.0004 & 0.01 & 0.9 & 1.4 \\
        \hline
        Brain  & 0.0178 & 1.25 & 0.9 & 1.4 \\
        \hline
        Blood  & 0.13   & 86.35 & 0.9835 & 1.33 \\
        \hline
    \end{tabular}
\end{table}

\subsection{Optode Configuration for Monte Carlo Extreme}

The configuration of sources and detectors, collectively known as optodes, is a yeah another important aspect of simulating light propagation in MCX simulations. Optode placement directly influences the spatial resolution and depth sensitivity of fNIRS measurements. For our MCX simulations, we mirror the optode configuration of our fNIRS device, which comprises 32 sources and 32 detectors arranged in a uniform grid.

The synthesized data aims to closely resemble that obtained from the fNIRS device used to capture Dataset 01 and 02. The device's optodes are arranged in a 2x16 grid format. In MCX, source and detector locations must be specified outside the scalp to simulate the actual conditions of an fNIRS measurement. These locations are derived from the fNIRS device's configurations and are adjusted to reside just outside the scalp in our parametric head model, ensuring an accurate representation of the measurement geometry.

The sources in our MCX simulations emit light in a 'pencil' manner, meaning they emit photons in a specific direction, which is typically perpendicular to the scalp surface at the point of emission. The directionality and intensity parameters affect the penetration depth and the area of the brain being illuminated.

Detectors in MCX are configured to capture photons exiting the scalp, registering their amount, exit locations and the total weight of the photon packet, which correlates with the amount of light absorbed and scattered through the brain tissue. The detectors, governed by their radius, is set to match the actual fNIRS detectors.

For our simulations, we have adapted the source and detector locations from our fNIRS device setup. The resultant model may be biased for the specific geometry. The spherical coordinates are converted to Cartesian coordinates to fit the MCX input requirements, ensuring that the sources and detectors are appropriately placed relative to our parametric head model. Table \ref{tab:sources_detectors} summarizes an exemplary subset of converted locations and launch directions for sources and detectors used in our simulations.

\begin{table}[t]
    \caption{Example locations and launch directions for sources and detectors in Monte Carlo simulations, based on the defined head model center and scalp radius.}
    \label{tab:sources_detectors}
    \centering
    \begin{tabular}{|l|c|c|}
        \hline
        \textbf{Optode} & \textbf{Location (mm)} & \textbf{Launch Direction} \\
        \hline
        \hline
        Detector 0 & (99.41, 184.58, 83.35) & N/A \\
        \hline
        Detector 1 & (99.41, 184.58, 102.65) & N/A \\
        \hline
        Detector 2 & (116.76, 181.67, 83.35) & N/A \\
        \hline
        Detector 3 & (116.76, 181.67, 102.65) & N/A \\
        \hline
        Detector 4 & (134.68, 174.79, 83.35) & N/A \\
        \hline
        Detector 5 & (134.68, 174.79, 102.65) & N/A \\
        \hline
        Source 0 & (89.77, 185.25, 93.0) & (3.23, -92.25, 0.0) \\
        \hline
        Source 1 & (108.74, 182.23, 110.62) & (-15.74, -89.23, -17.62) \\
        \hline
        Source 2 & (109.03, 183.9, 93.0) & (-16.03, -90.9, 0.0) \\
        \hline
        Source 3 & (125.47, 177.59, 110.62) & (-32.47, -84.59, -17.62) \\
        \hline
        Source 4 & (126.08, 179.17, 93.0) & (-33.08, -86.17, 0.0) \\
        \hline
        Source 5 & (142.35, 168.99, 110.62) & (-49.35, -75.99, -17.62) \\
        \hline
        \multicolumn{3}{|p{8cm}|}{Note: This table provides an example of the locations and launch directions for sources and detectors used in Monte Carlo simulations for fNIRS. The positions are calculated based on a centered head model with a center at (93, 93, 93) and a scalp radius of 92.3 mm. The angular positions for the detectors and sources have been converted to Cartesian coordinates considering these assumptions. This representation illustrates a subset of the used optodes to showcase the configuration setup for Monte Carlo simulations.} \\
        \hline
    \end{tabular}
\end{table}

\subsection{Simulation Parameters}
The choice of simulation parameters balance computational efficiency with the accuracy required for generating meaningful synthetic data. Apart from the number of photons, we did not change the default parameters provided by MCX. Preliminary tests suggested that the default settings offered a good balance between time and resolution. Fine-tuning these parameters may yield different insights into the model's behavior and potentially enhance the accuracy or efficiency of the simulations. 

Generally speaking, the more high-quality data a model has to train on, the better the model behaves. Given the scope of our project, performing tens of thousands of simulations, the number of photons for each simulation was set to $10^8$.  With each approximately $5 s$ in duration, the overall computational time posed a significant consideration. The chosen photon number ensures statistical robustness while keeping individual simulation times within a feasible range. 

\subsection{Understanding Monte Carlo eXtreme Simulation Outputs}

MCX simulations produce two primary types of output files: .jnii and .jdat, each serving a unique purpose in the context of photon migration and detection analysis within simulated tissues.

The .jnii files, an abbreviation for "JSON NIfTI," encapsulate the spatially resolved flux data. This format provides a comprehensive depiction of the distribution and intensity of light as it propagates through the simulated tissue environment. The .jnii files record the scattering and absorption events, thereby recording the spatial variations attributable to differing tissue properties. This mapping enables visualization of light propagation and interaction with various tissue components. This visualization can provide insight regarding photonic behavior within structures.

The .jdat files hold time-resolved information regarding photon detection. These files log the count of photons reaching each detector within the model, segmenting the data temporally to reflect the dynamics of photon capture. This temporal resolution is critical for the analysis of the light's time of flight. 

In the context of machine learning applications for fNIRS tomography, the .jdat file's detector-specific photon counts are particularly valuable. By extracting this data photon counts can be incorporated into a dataset representative of observed physiological conditions. This dataset then acts as the input for machine learning models, which are tasked with predicting the ground truth of the head model, focusing notably on the location and characteristics of simulated blood vessels or anomalies.

The potential utility of .jnii files in the context of machine learning, especially for backpropagation algorithms, should not be overlooked. These files could provide a spatial context for the photon counts recorded in .jdat files, offering a detailed background against which the machine learning models can refine their predictions. For instance, by comparing the spatial flux distributions from .jnii files with the photon count data, models may better learn how light propagates through tissue, enabling more accurate reconstructions of the underlying tissue structure. It is important to mention that flux measurements are only available for simulations. Real world recordings of fNIRS experiments yield photon counts at detector locations, thus the motivation to extract this data from the simulated environment.

\begin{figure}[!t]
    \centering
    \includegraphics[width=\columnwidth, keepaspectratio]{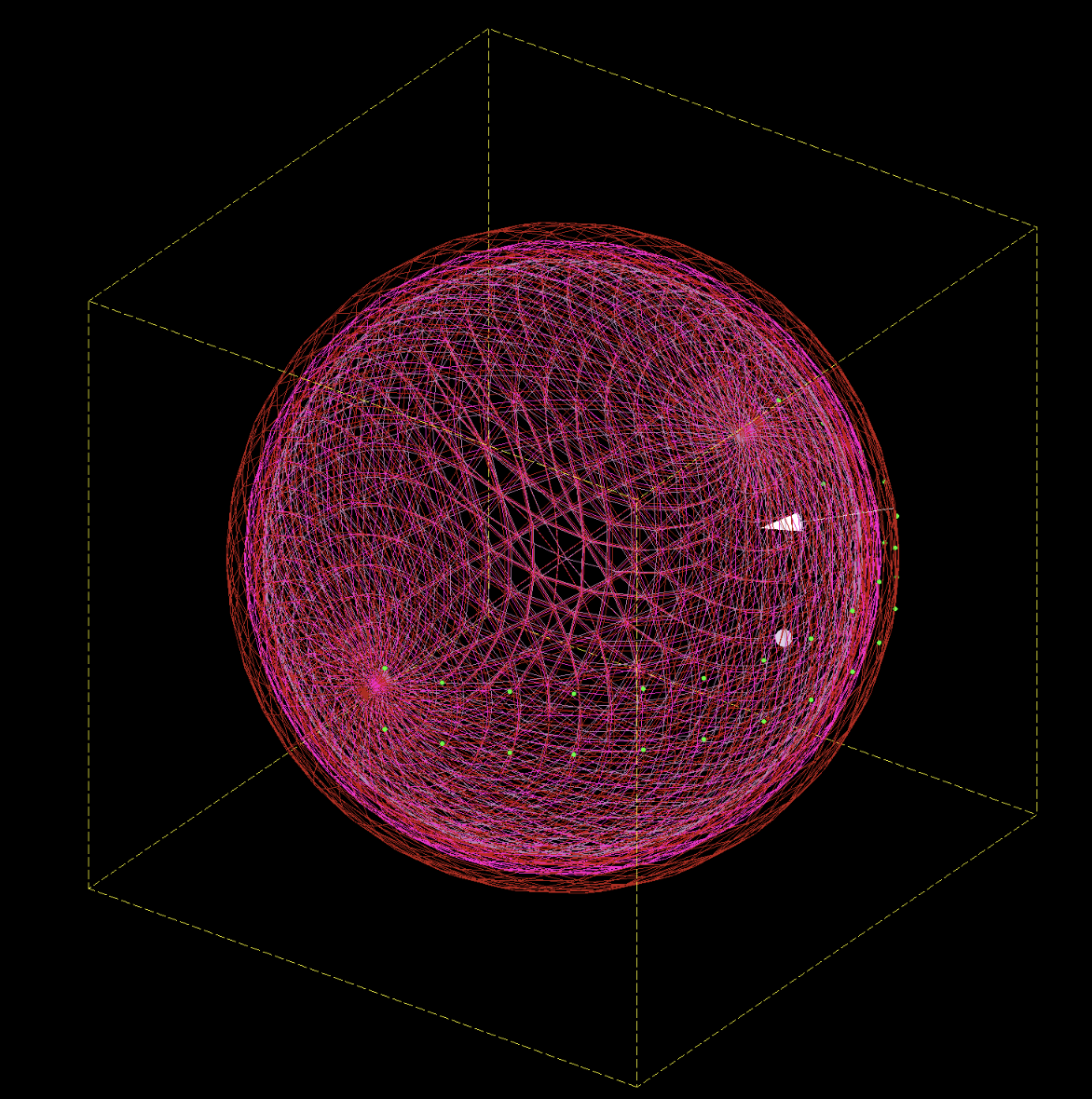}
    \caption{Isometric view of Monte Carlo eXtreme simulation, illustrating the layered concentric spheres representing different head tissues. The outermost sphere represents the scalp, with detectors (green) placed around its surface. The white inner sphere represents a blood region of interest. A directional arrow indicates the photon emission source, aimed perpendicular to the scalp's surface, demonstrating the initial photon trajectory towards the tissue layers.}
    \label{fig:mcx_preview_01}
\end{figure}

\begin{figure}[!t]
    \centering
    \includegraphics[width=\columnwidth, keepaspectratio]{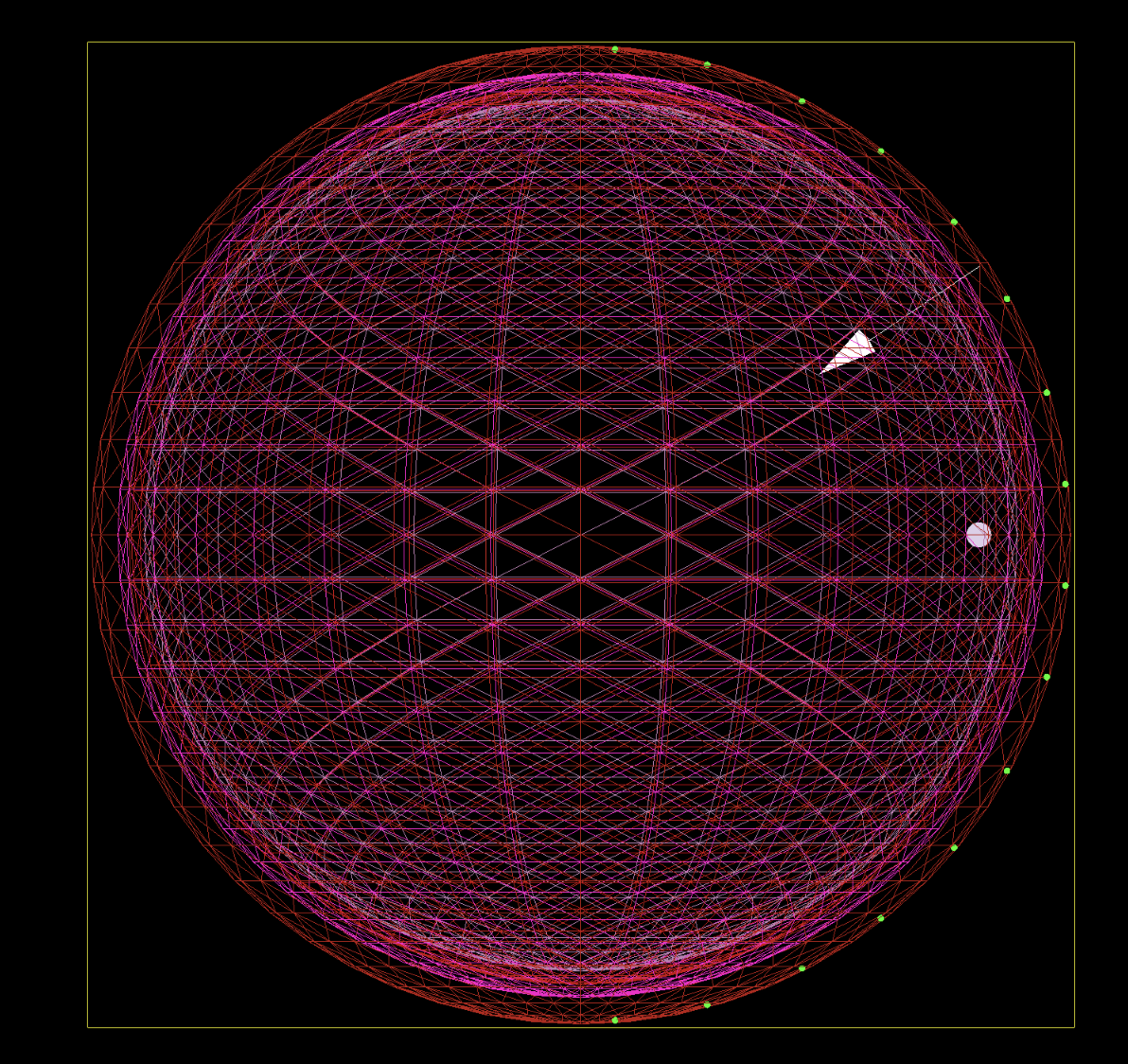}
    \caption{Top view of the Monte Carlo eXtreme simulation. This figure showcases the geometric arrangement of detectors (green) around the scalp's simulation sphere. It provides a clear view of the spatial relationship between the detectors and the embedded white sphere representing a target blood volume within the head's tissue layers.}
    \label{fig:mcx_preview_02}
\end{figure}

\begin{figure}[!t]
    \centering
    \includegraphics[width=\columnwidth, keepaspectratio]{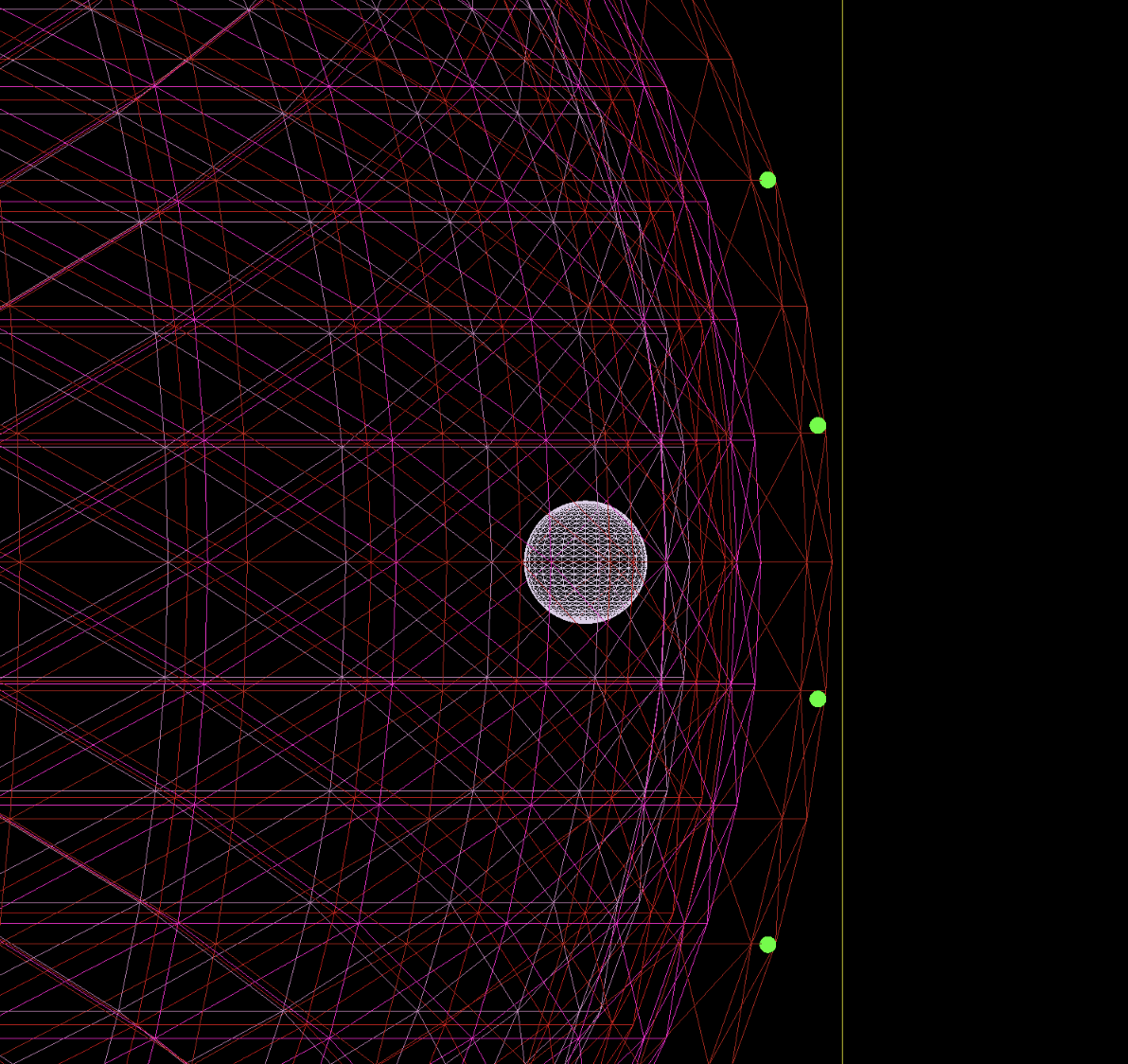}
    \caption{Close-up view of the Monte Carlo eXtreme simulation focusing on the region near the blood-representing sphere. This figure illustrates the concentric tissue layers and the positioning of the detectors.}
    \label{fig:mcx_preview_03}
\end{figure}

\subsection{Data Generation}

To efficiently execute vast amounts of MCX simulations, a structured automation system is desired. Leveraging the advantages of an SQL database to ensure organized data storage, streamlined data retrieval, and modification, allow for the automation of generating JSON configuration files. Each MCX simulation requires an input JSON file containing parameters that define a given simulation, such as geometries and their respective optical properties, optode types and locations, and the number of photons to simulate. A MySQL database was engineered to mirror and extend MCX's simulation parameters, incorporating specialized tables for detailed configuration. The database is utilized to systematically generate and manage these JSON files, each representing a unique simulation scenario, thereby enhancing the efficiency and scalability of the simulation process.

\begin{figure}[!t]
    \centering
    \includegraphics[width=\columnwidth, keepaspectratio]{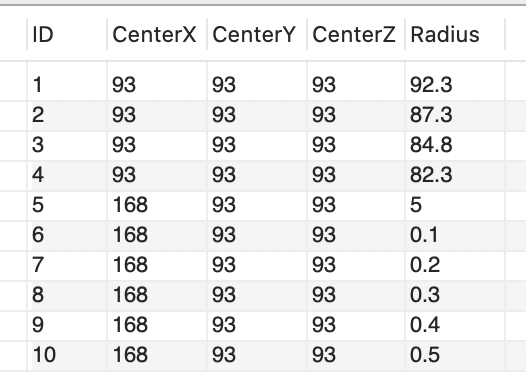}
    \caption{The sphere table from the MySQL database, detailing spherical objects utilized in the Monte Carlo simulations. Columns include ID for unique identification, CenterX, CenterY, CenterZ for center coordinates of each sphere, and Radius for the size of each sphere. This table supports the creation of parametric models in simulations, allowing for the variation of sphere locations and sizes.}
    \label{fig:Sphere_Table}
\end{figure}

\begin{figure}[!t]
    \centering
    \includegraphics[width=\columnwidth, keepaspectratio]{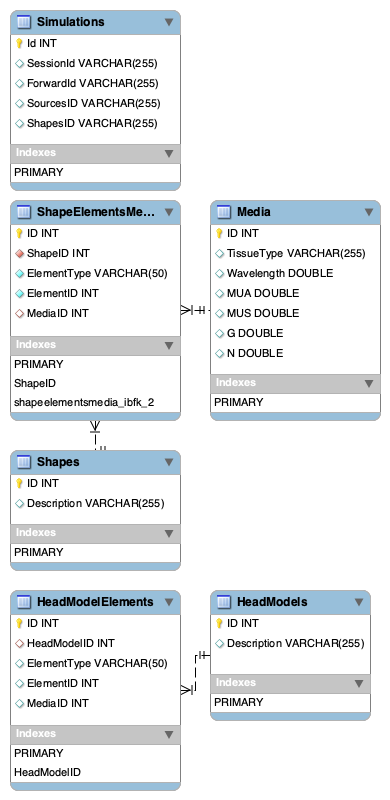}
    \caption{Part of the MySQL database schema relevant to the synthetic data generation for fNIRS simulations. This partial schema shows tables that do not directly mirror Monte Carlo eXtreme (MCX) parameters but are essential for defining the configurations of simulations. This includes Simulations, ShapeElementsMedia, Media, Shapes, HeadModelElements, and HeadModels, illustrating their relationships and data structures necessary for generating MCX input files.}
    \label{fig:partial_schema}
\end{figure}

Each record in the 'Simulations' table describes a simulation scenario. The columns within this table - SessionId, ForwardId, SourcesID, and ShapesID — establish links to corresponding tables, each containing the associated configuration options for MCX. 

The 'Shapes' table does not map to an MCX parameter. It orchestrates the shapes associated with each simulation. The 'ShapeElementsMedia' table brings together shape elements and media, interlinking shapes with their respective optical properties, thereby forming the structure used in MCX simulations. Each shape element is associated with its relevant parameters and media properties, facilitating the generation of a detailed and varied synthetic dataset.

Figures \ref{fig:Sphere_Table} and \ref{fig:partial_schema} illustrate the several components of our MySQL database. Figure \ref{fig:Sphere_Table} depicts the `Sphere` table, exemplary in defining shape parameters in simulations, this specific geometry having parameters such as sphere sizes and locations. Figure \ref{fig:partial_schema} presents a section of the database schema highlighting the interconnected tables designed to extend beyond the basic MCX configuration parameters.

Within our constructed head model for synthetic data generation, the initial four records of the `Sphere` table play define the head's geometrical basis, as illustrated in Figures \ref{fig:Sphere_Table} and \ref{fig:mcx_preview_03}. These records correspond to concentric spheres representing the layers of the head, such as the scalp, skull, cerebrospinal fluid, and the brain itself. This parametric head model, utilizing simplified geometry, underpins our simulations, approximating the biological structures of the human head. By embedding various geometric elements within this parametric model, with optical properties of blood, we can simulate many different scenarios.

To test the feasibility of this approach we have initially focused on scenarios involving only spherical elements. The first of these has a static center position for the sphere, while varying its radius from 0.1 mm to 7.2 mm in increments of 0.1 mm, reflecting variances in pseudo blood vessel size. The second scenario maintains a constant sphere radius of 7 mm but alters the sphere's central position from -10° to 10° in elevation and from -180° to 180° in azimuth in 1° increments, this time varying pseudo blood vessel location. 

The variability in pseudo blood vessel locations and sizes is essential for ensuring that our machine learning models are exposed to a wide range of states. This diversity aids in enhancing the models' generalizability.

\subsection{Generating Simulation Configuration Files}

The process of generating simulation configuration files is automated through a Python script, which interacts with the MySQL database to fetch simulation parameters and then constructs the necessary JSON files. Here, we explain the methodology and algorithms employed by the script to automate this task.

The Python script employs a class \texttt{SimulationSQLHandler} which establishes a connection to the database, fetches simulation parameters, and generates JSON files for each simulation setup.

\begin{algorithm}
\caption{Generating JSON files}
\begin{algorithmic}[1] 
\STATE Establish database connection using MySQL connector based on credentials.
\STATE Fetch all simulation IDs from the \texttt{Simulations} table.
\FOR{each simulation ID}
    \STATE Fetch associated session, forward model, shapes, and sources data.
    \STATE Construct a unique session ID and determine if a JSON file already exists.
    \IF{JSON file does not exist}
        \STATE Fetch shape elements and media from the database.
        \STATE Process and structure data into a JSON-compatible format.
        \STATE Adjust angles and shapes based on simulation specifics.
        \STATE Combine source and detector configurations into the optode structure.
        \STATE Save the constructed data into a JSON file.
    \ENDIF
\ENDFOR
\STATE Close database connection.
\end{algorithmic}
\end{algorithm}

The script starts by fetching all necessary data for the simulations from various tables within the database, such as \texttt{Session}, \texttt{Forward}, \texttt{Shapes}, and \texttt{Sources}. Each table contains parameters for the MCX simulation setup, including the geometrical models of the head, light source positions, and detector locations.

After fetching the data, the script processes this information to comply with MCX requirements. This involves translating spherical coordinates into Cartesian for source and detector positions, adjusting angles, and setting up the simulation domain. The script handles different shape structures, particularly spheres, by adjusting their positions based on simulation specifics.

Furthermore, the Python script constructs the JSON file, structuring it to reflect the session, forward model, optodes, shapes, and domain configurations. This file is then used directly by MCX to initiate the simulations.

This automation significantly reduces the time and effort required to set up simulations, enabling the execution of extensive datasets necessary for training machine learning models. By systematically generating these configurations, we ensure consistency and accuracy across all simulated scenarios.

The Python script's interaction with the MySQL database is designed to prevent errors and ensure efficient data handling. The connection to the database is securely established, and data retrieval is conducted with error checks to avoid interruptions during the script's execution. Once the JSON files are generated and saved, the script closes the database connection to maintain data integrity and system security.

\subsection{Setting Up and Running Simulations in Parallel on AWS}
The number of MCX simulations that can run in parallel on a single machine is restricted by the machine's hardware capabilities, particularly in terms of GPU and CPU resources. Cloud computing addresses this bottleneck by enabling scalable computational resources that can be adjusted according to the workload's demands. Beyond scalability, cloud computing ensures a standardized computational environment, essential for reproducibility and consistency across simulations. Additionally, it efficiently handles and stores large datasets. Accordingly, we established an AWS infrastructure designed to streamline our fNIRS simulation processes, optimizing resource utilization, and reducing overall computation time.

Utilizing AWS CloudFormation, we established a scalable and secure cloud infrastructure to support extensive Monte Carlo simulations for fNIRS data analysis. This setup orchestrates components through the CloudFormation template, ensuring a reproducible and systematic environment for data generation and analysis.

Our AWS setup is composed of components designed to establish a secure computational environment. A Virtual Private Cloud (VPC) coupled with an Internet Gateway create a secure network environment that facilitates controlled access to computational resources while enabling internet connectivity for updates and data transfers. Within this VPC, subnets, route tables, and security groups manage and secure traffic.

An S3 Bucket has been set up for structured data storage. The S3 bucket is divided into two main directories: mcx\_input\_jsons and mcx\_output. The mcx\_input\_jsons directory serves as a repository for storing the input JSON files necessary for initiating the MCX simulations, while the mcx\_output directory is designated for storing the resulting output files from these simulations. 

We utilize G type EC2 instances equipped with Nvidia GPUs. The initial setup is performed on a `g4dn.xlarge` instance. Our EC2 instances are configured for running Monte Carlo simulations. Initially, we update all system packages and install the necessary tools to ensure that the operating environment is up-to-date and secure. Following this, we set up the CUDA Toolkit, an MCX requirement essential for enabling GPU-powered processing. We then implement AWS CLI v2 for AWS resource management. Later we configure environment variables and directory structures to support MCX simulations. We automate the synchronization of data with the S3 bucket. This automation enables efficient handling of simulation inputs and outputs. 

With this setup, our EC2 instances become capable of executing Monte Carlo simulations in parallel.

Our operational framework uses the script \texttt{run\_simulations\_2024.2.sh} for managing simulation executions and data transfers between EC2 instances and the S3 bucket.

The architecture is designed for scalability, allowing dynamic allocation of computational resources to accommodate the workload. We plan to enhance our system by implementing AWS Batch for parallel processing across multiple EC2 instances, further reducing computation time.

Security measures include security groups and IAM policies, limiting access to necessary services and ports while preventing unauthorized data access. Our S3 buckets are secured with access permissions, ensuring data integrity and confidentiality.

Deployment of our AWS-based computational framework facilitates the generation of large-scale synthetic datasets essential for machine learning applications. This cloud-based approach not only improves the efficiency and scalability of data generation but also ensures the reproducibility and security of the computational environment.

\section{Conclusion}
This article provides an overview of functional near-infrared spectroscopy (fNIRS) as an established method for studying neurovascular phenomena. The article proposes a systematic approach for setting up a data environment to enable meaningful comparisons between different signal processing and machine learning modalities. To facilitate this, the authors set up a containerized environment using industry-standard tools for data provenance and consistency, which lays the infrastructure for subsequent research towards blind tomography and semantic-hemodynamic space transformation. The article also provides a brief history of the development of optical methods for assessing changes in the optical properties of brain tissue and discusses the physics of fNIRS, including the Beer-Lambert Law and the Modified Beer-Lambert Law, as well as the cortical hemodynamic response to brain activity. Further, the article briefly describes the components of fNIRS systems and the factors that influence the quality of measurement data.

The datasets were stored in the NetCDF file format, which is a binary file format used to store scientific data and metadata in a self-describing form. The Xarray library was used to load the datasets, which provides support for labeled, multi-dimensional arrays and integrates with a range of other scientific Python packages. The Docker platform was used to create a self-contained environment for running the Jupyter notebook with the necessary dependencies for fNIRS. The Dockerfile was used to build an image for a Jupyter notebook with the required dependencies, and the mamba package manager was used to install additional dependencies specified in the requirements.txt file. The study's data analysis environment was thus standardized and reproducible, which enhances data provenance and simplifies the installation and configuration of software and its dependencies.

Using a novel fd-fNIRS device two datasets were recorded. Dataset 01 consists of three experiment runs, and provides detailed information on its dimensions, coordinates, and values. The dataset is complex in nature, capturing both the magnitude and phase components of the fd-fNIRS system. Spatial information is provided for each detector and source in three dimensions, and a function is used to estimate the z values based on the subject's head circumference. The article highlights the importance of SDS and NN values in determining the probability of detecting photons, and how the difference between two-dimensional and three-dimensional distances between the sources and detectors may have a significant impact on DOT calculations. The author found that detecting faulty sources or detectors can be done by comparing the mean of magnitude data across time with the average value for source detector pairs with the same NN value, and visualizing them in a heat map to detect coupling strength. 

The concept of semantic representation is introduced as the process of capturing the meaning of natural language text in a way that can be used to reason about, understand, and manipulate its content. The use of fMRI imaging for predicting nouns is discussed, along with the advantages of using fNIRS imaging over fMRI. The article then presents Dataset 02, which explores the relationship between semantic representation and neurovascular coupling data obtained using fNIRS, including its dimensions and variables. The article also provides a list of questions that can be used to classify objects into different categories based on their appearance, function, and other characteristics.

Presumably, knowledge of spatio-temporal cortical hemodynamics could allow for finding correlation between said hemodynamics and different states of the brain, whether pathological or consciousness related, akin to research performed with different imaging modalities such as fMRI.

This research has demonstrated the potential of synthetic data, generated through Monte Carlo simulations, in addressing the challenge of limited high-quality datasets in neuroimaging, particularly in functional near-infrared spectroscopy (fNIRS) tomography. The approach detailed within allows for the controlled exploration of a wide array of conditions, thus facilitating the training and validation of machine learning models specifically designed for fNIRS data analysis.

The employment of parametric head models, alongside photon propagation simulation techniques, provide for generating diverse datasets. With proper simulation configurations, these datasets may not only encompass a broad spectrum of variables but may also reflect realistic physiological conditions, thereby laying a solid foundation for enhancing machine learning applications within fNIRS tomography. Furthermore, this study has established a comprehensive framework that includes the use of data storage in NetCDF format, efficient data handling through Xarray, and the strategic deployment of Docker containers. These methodologies collectively foster systematic and reproducible data analysis, a crucial factor in advancing neuroimaging research.

Moreover, the establishment of a cloud-based infrastructure promote scalability and accessibility of high-quality neuroimaging data. This infrastructure, designed for efficient data generation and processing, not only supports the creation of extensive datasets imperative for the development of robust machine learning models but also ensures the consistency and reproducibility essential for credible scientific studies.

The successful application of machine learning techniques to fNIRS data relies heavily on the availability of such extensive, high-quality datasets. By providing a reliable solution to data scarcity and setting new standards in data simulation and analysis, this work paves the way for future research aimed at advancing the accuracy, efficiency, and overall applicability of fNIRS tomography. The resultant potential for improved diagnostic tools and treatment strategies opens new avenues for addressing neurological conditions, promising significantly enhanced patient outcomes and deeper insights into brain dynamics.

In conclusion, the advancements introduced in this study highlight the transformative power of integrating machine learning with synthetic data generation in the field of neuroimaging. The developed methodologies not only serve as a comprehensive blueprint for future research but also contribute significantly to the progression of fNIRS tomography, thereby making a notable impact on the broader domain of biomedical engineering.

\appendices
\section{JupyterLab Code}
The code can be accessed at the following GitHub repository:

\url{https://github.com/EitanWaks/Independent-Study.git}

\section{Feature Description Questions}
\label{appendix_questions}
The complete list of questions for the coordinate \texttt{feature\_desc} are:
\begin{itemize}
    \item Is it an animal? 
    \item You what's up me Is it a body part? 
    \item Is it a building? 
    \item Is it a building part? 
    \item Is a clothing?
    \item Is it furniture?
    \item Is it an insect?
    \item Is it a kitchen item?
    \item Is it man-made?
    \item Is it a tool?
    \item Can you eat it?
    \item Is it a vehicle?
    \item Is it a person?
    \item Is it a vegetable/plant?
    \item Is it a fruit?
    \item Is it made of metal?
    \item Is it made of plastic?
    \item Is part of it made of glass?
    \item Is it made of wood?
    \item Is it shiny?
    \item Can you see through it?
    \item Is it colorful?
    \item Does it change color?
    \item Is it more than one colored?
    \item Is it always the same color(s)?
    \item Is it white?
    \item Is it red?
    \item Is it orange?
    \item Is it flesh colored?
    \item Is it yellow?
    \item Is it green?
    \item Is it blue?
    \item Is it silver?
    \item Is it brown?
    \item Is it black?
    \item Is it curved?
    \item Is it straight?
    \item Is it flat?
    \item Does it have a front and back?
    \item Does it have a flat/straight top?
    \item Does it have flat/straight sides?
    \item Is it taller than it is wide/long?
    \item Is it long?
    \item Is it pointed/sharp?
    \item Is it tapered?
    \item Is it round?
    \item Does it have corners?
    \item Is it symmetrical?
    \item Is it hairy?
    \item Is it fuzzy?
    \item Is it clear?
    \item Is it smooth?
    \item Is it soft?
    \item Is it heavy?
    \item Is it lightweight?
    \item Is it dense?
    \item Is it slippery?
    \item Can it change shape?
    \item Can it bend?
    \item Can it stretch?
    \item Can it break?
    \item Is it fragile?
    \item Does it have parts?
    \item Does it have moving parts?
    \item Does it come in pairs?
    \item Does it come in a bunch/pack?
    \item Does it live in groups?
    \item Is it part of something larger?
    \item Does it contain something else?
    \item Does it have an internal structure?
    \item Does it open?
    \item Is it hollow?
    \item Does it have a hard inside?
    \item Does it have a hard outer shell?
    \item Does it have at least one hole?
    \item Is it alive?
    \item Was it ever alive?
    \item Is it a specific gender?
    \item Is it manufactured?
    \item Was it invented?
    \item Was it around 100 years ago?
    \item Are there many varieties of it?
    \item Does it come in different sizes?
    \item Does it grow?
    \item Is it smaller than a golf ball?
    \item Is it bigger than a loaf of bread?
    \item Is it bigger than a microwave oven?
    \item Is it bigger than a bed?
    \item Is a parrot in a car?
    \item Is it bigger than a house?
    \item Is it taller than a person?
    \item Does it have a tail?
    \item Does it have legs?
    \item Does it have four legs?
    \item Does it have feet?
    \item Does it have paws?
    \item Does it have claws?
    \item That I have horns/thorns/spikes?
    \item Does it have hooves?
    \item Does it have a face?
    \item Does it have a backbone?
    \item Does it have wings?
    \item Does it have ears?
    \item Does it have roots?
    \item Does it have seeds?
    \item Does it have leaves?
    \item Does it come from a plant?
    \item Does it have feathers?
    \item Does it have some sort of nose?
    \item Does it have a hard nose/beak?
    \item Does it contained liquid?
    \item Does it have wires or record?
    \item Does it have writing on it?
    \item Does it have wheels?
    \item Does it make a sound?
    \item Does it make a nice sound?
    \item Does it make a sound continuously when active?
    \item Is it job to make sound?
    \item Does it roll?
    \item Can it run?
    \item Is it fast?
    \item Can it fly?
    \item Can it jump?
    \item Can it float?
    \item Can it swim?
    \item Can it dig?
    \item Can it climb trees?
    \item Can it cause you pain?
    \item Can it bite or sting?
    \item Does it stand on two legs?
    \item Is it wild?
    \item Is it a herbivore?
    \item Is it a predator?
    \item Is it warm blooded?
    \item Is it a mammal?
    \item Is it nocturnal?
    \item Does it lay eggs?
    \item Is it conscious?
    \item Does it have feelings?
    \item Is it smart?
    \item Is it mechanical?
    \item Is it electronic?
    \item Does it use electricity?
    \item Can it keep you dry?
    \item Does it provide protection?
    \item Does it provide shade?
    \item Does it cast a shadow?
    \item Do you see it daily?
    \item Is it helpful?
    \item Do you interact with it?
    \item Can you touch it?
    \item Would you avoid touching it?
    \item Can you hold it?
    \item Can you hold it in one hand?
    \item Do you hold it to use it?
    \item Can you play it?
    \item Can you play with it?
    \item Can you pet it?
    \item Can you use it?
    \item Do you use it daily?
    \item Can you use it up?
    \item Do you use it when cooking?
    \item Is it used to carry things?
    \item Can you pick it up?
    \item Can you control it?
    \item Can you sit on it?
    \item Can you ride on/in it?
    \item Is it used for transportation?
    \item Can you fit inside it?
    \item Is it used in sports?
    \item Do you wear it?
    \item Can it be washed?
    \item Is it cold?
    \item Is it cool?
    \item Is it warm?
    \item Is it hot?
    \item Is it unhealthy?
    \item Is it hard to catch?
    \item Can you peel it?
    \item Can you walk on it?
    \item Can you switch it on and off?
    \item Can it be easily moved?
    \item Do you drink from it?
    \item Does it go in your mouth?
    \item Is it tasty?
    \item Is it used during meals?
    \item Does it have a strong smell?
    \item Does it smell good?
    \item Does it smell bad?
    \item Is it usually inside?
    \item Is it usually outside?
    \item Would you find it on a farm?
    \item Would you find it in a school?
    \item Would you find it in a zoo?
    \item Would you find it in an office?
    \item Would you find it in a restaurant?
    \item Would you find it in the bathroom?
    \item Would you find it in a house?
    \item Would you find it near a road?
    \item Would you find it in a dump/landfill?
    \item Would you find it in the forest?
    \item Would you find it in a garden?
    \item Would you find it in the sky?
    \item And do you find it in space?
    \item Does it live above ground?
    \item Does it get wet?
    \item Does it live in water?
    \item Can it live out of water?
    \item Do you take care of it?
    \item Does it make you happy?
    \item Do you love it?
    \item Would you miss it if it were gone?
    \item Is it scary?
    \item Is it dangerous?
    \item Is it friendly?
    \item Is it rare?
    \item Can you buy it?
    \item Is it valuable?
\end{itemize}

\section*{Acknowledgment}

The author would like to thank Griffin Millsap, Brock Wester, and Anil Maybhate for their support, patience, guidance and goodwill.

\newpage

\printbibliography

\begin{IEEEbiography}[{\includegraphics[width=1in,height=1.25in,clip,keepaspectratio]{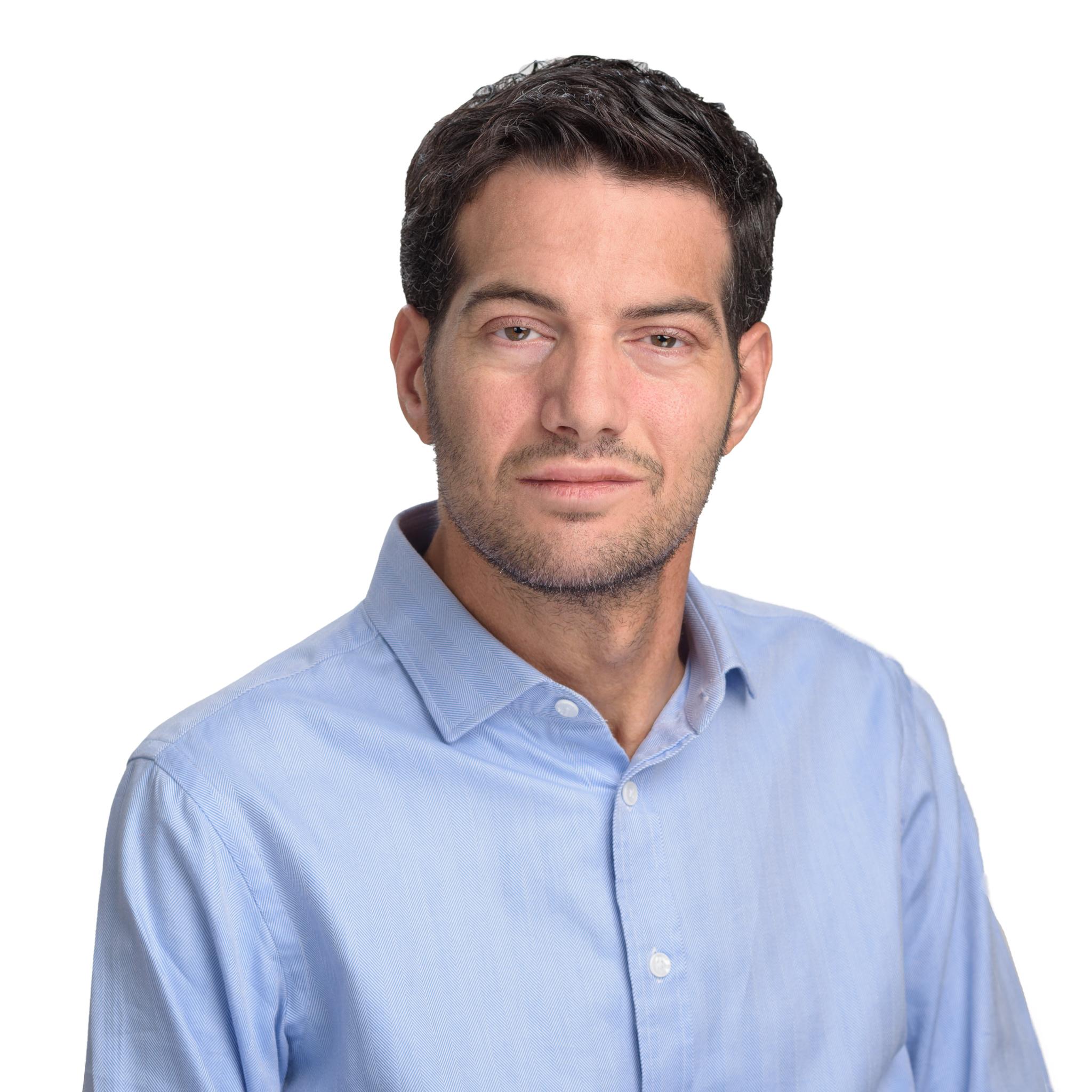}}]{Eitan Waks}
studied mechanical engineering in the Technion, graduating with a bachelors degree in 2009. After several years in industry, he became a registered US Patent Agent in December 2013. Shortly there after he founded E. Waks \& Co. After leaving the company in 2021, he began a masters program in Applied Biomedical Engineering at Johns Hopkins University. His research focus is in neuro-engineering, primarily novel neuroimaging techniques and brain computer interfaces. 

\end{IEEEbiography}

\end{document}